\documentclass[a4paper,twocolumn,11pt,accepted=2023-03-11]{quantumarticle}
\pdfoutput=1
\usepackage[utf8]{inputenc}
\usepackage[english]{babel}
\usepackage[T1]{fontenc}
\usepackage[numbers]{natbib}
\usepackage{listings}

\usepackage{amsmath}
\usepackage{amssymb}
\usepackage{hyperref}
\usepackage[resetlabels]{multibib}

\newcites{article}{article references}
\newcites{book}{book references}
\newcites{misc}{misc references}
\newcites{repo}{repository references}
\newcites{web}{website references}
\newcites{other}{Other references}
\pdfoutput=1
\usepackage[utf8]{inputenc}
\usepackage[english]{babel}
\usepackage[T1]{fontenc}
\usepackage{amsmath}
\usepackage{hyperref}
\usepackage{tikz}
\usepackage{lipsum}
\usepackage{graphicx,amsmath,amsfonts,amssymb,amsthm,xr}
\usepackage{newpxtext}
\usepackage{epsfig,amsmath,amssymb,color,dsfont,upgreek,physics}
\usepackage{mathrsfs}
\usepackage{mathtools}
\usepackage{bbold}

\definecolor{mygold}{rgb}{0.93,0.69,0.13}

\definecolor{mypurple}{rgb}{0.49,0.18,0.56}

\definecolor{mygreen}{rgb}{0,0.5,0}

\definecolor{mygreen}{rgb}{0,0.5,0}
\definecolor{myred}{rgb}{0.7,0,0}

\begin{document}
\title{Robust quantum many-body scars in lattice gauge theories}
\author{Jad C.~Halimeh}
\email{jad.halimeh@physik.lmu.de}
\affiliation{Department of Physics and Arnold Sommerfeld Center for Theoretical Physics (ASC), Ludwig-Maximilians-Universit\"at M\"unchen, Theresienstra\ss e 37, D-80333 M\"unchen, Germany}
\affiliation{Munich Center for Quantum Science and Technology (MCQST), Schellingstra\ss e 4, D-80799 M\"unchen, Germany}
\author{Luca Barbiero}
\affiliation{Institute for Condensed Matter Physics and Complex Systems, DISAT, Politecnico di Torino, I-10129 Torino, Italy}
\author{Philipp Hauke}
\affiliation{INO-CNR BEC Center and Department of Physics, University of Trento, Via Sommarive 14, I-38123 Trento, Italy}
\author{Fabian Grusdt}
\affiliation{Department of Physics and Arnold Sommerfeld Center for Theoretical Physics (ASC), Ludwig-Maximilians-Universit\"at M\"unchen, Theresienstra\ss e 37, D-80333 M\"unchen, Germany}
\affiliation{Munich Center for Quantum Science and Technology (MCQST), Schellingstra\ss e 4, D-80799 M\"unchen, Germany}
\author{Annabelle Bohrdt}
\affiliation{ITAMP, Harvard-Smithsonian Center for Astrophysics, Cambridge, MA 02138, USA}
\affiliation{Department of Physics, Harvard University, Cambridge, Massachusetts 02138, USA}

\begin{abstract}
Quantum many-body scarring is a paradigm of weak ergodicity breaking arising due to the presence of special nonthermal many-body eigenstates that possess low entanglement entropy, are equally spaced in energy, and concentrate in certain parts of the Hilbert space. Though scars have been shown to be intimately connected to gauge theories, their stability in such experimentally relevant models is still an open question, and it is generally considered that they exist only under fine-tuned conditions. In this work, we show through Krylov-based time-evolution methods how quantum many-body scars can be made robust in the presence of experimental errors through utilizing terms linear in the gauge-symmetry generator or a simplified pseudogenerator in $\mathrm{U}(1)$ and $\mathbb{Z}_2$ lattice gauge theories. Our findings are explained by the concept of quantum Zeno dynamics. Our experimentally feasible methods can be readily implemented in existing large-scale ultracold-atom quantum simulators and setups of Rydberg atoms with optical tweezers.
\end{abstract}

\maketitle

\section{Introduction}
The thermalization of closed quantum systems under unitary time evolution \cite{Deutsch1991,Srednicki1994} has intrigued physicists for decades due to the counterintuitive picture of thermalization in the absence of a reservoir. Even though a pure quantum state has zero entropy throughout its entire unitary time evolution, quantum entanglement \cite{Calabrese2005,Amico2008,Daley2012,Schachenmayer2013} can be a guiding principle in understanding the ``thermalization'' of its constituent subsystems based on the concepts of statistical mechanics. The entropy of entanglement between different subsystems can cause them to equilibrate, and for generic nonintegrable models and typical initial states, local observables can then be described by thermal ensembles in the long-time limit in accordance with the eigenstate thermalization hypothesis (ETH) \cite{Rigol_2008,Eisert2015,Rigol_review,Deutsch_review}. This has been demonstrated experimentally in a Bose--Hubbard optical lattice~\cite{Kaufmann2016}.

However, it has been shown that ETH can be violated in several ways. When the system is integrable \cite{sutherland2004beautiful}, for example, it has infinitely many conserved local integrals of motion (LIOMs), which prevent it from thermalizing. Instead, it equilibrates to a generalized Gibbs ensemble \cite{Rigol2007,Vidmar_review}. Closely related to integrability, interacting systems with quenched disorder can also violate ETH and exhibit many-body localization (MBL) \cite{Basko2006,Nandkishore_review,Alet_review,Abanin_review}, with emergent LIOMs that prevent thermalization. Quenched-disorder MBL has also been demonstrated experimentally in ultracold-atom setups \cite{Smith2016,Choi2016}. And more recently, it has been shown that even disorder-free interacting spin models can exhibit so-called Stark MBL in the presence of a constant magnetic field \cite{Schulz2019}, with emergent \textit{dynamical} quasi LIOMs \cite{gunawardana2022dynamical}. Stark MBL has also been experimentally observed in a trapped-ion quantum simulator \cite{Morong2021}.

Interestingly, many-body localization can still occur in disorder-free spatially homogeneous quantum many-body models, even when the latter are nonintegrable \cite{smith2017,Brenes2018,Metavitsiadis2017,smith2017absence,Smith2018,Russomanno2020,Papaefstathiou2020,karpov2021disorder,hart2021logarithmic,Zhu2021,Sous2021,Chakraborty2022}. This usually occurs when the underlying model hosts a local gauge symmetry, and the initial state is prepared in a superposition of an extensive number of gauge superselection sectors. Upon quenching such an initial state, \textit{disorder-free localization} (DFL) can arise, where the initial state's superposition over the symmetry sectors dynamically generates an effective disorder over the corresponding \textit{background charges} of these gauge sectors. DFL has been shown to vanish in the presence of errors breaking the local symmetry \cite{Smith2018}, but recent theoretical works have demonstrated that DFL can be stabilized or even enhanced using linear gauge protection schemes that we will discuss and employ in this work \cite{Halimeh2021stabilizingDFL,Halimeh2021enhancing,lang2022disorder}.

The above paradigms of strong ergodicity breaking are also complemented by a different form of weak ergodicity breaking occurring in nonintegrable models hosting \textit{quantum many-body scars} \cite{Turner2018,Serbyn2020,MoudgalyaReview,Zhao2021}. These are special nonthermal eigenstates characterized by anomalously low entanglement entropy, although they can be in the bulk of the system far away from the ground state. They also are roughly equally spaced in energy, and cluster in a \textit{cold subspace} weakly connected to the rest of the Hilbert space \cite{ShiraishiMori,Moudgalya2018,BernevigEnt,lin2018exact,Iadecola2019_2,MotrunichTowers}. As such, preparing an initial state in this cold subspace and then quenching it will result in scarred dynamics in the form of persistent oscillations beyond local relaxation timescales, significantly delaying the onset of thermalization \cite{Bernien2017,Zhao2020,Su2022}. It is important to note that quantum many-body scars are not directly connected to any underlying symmetry of the model, and, in contrast to DFL, a superposition over different symmetry sectors is not required for the emergence of nonthermal dynamics.

The Rydberg-atom setup \cite{Bernien2017}, in which quantum many-body scarring was first observed, implemented an Ising-type spin model that has since been shown to map onto the spin-$1/2$ $\mathrm{U}(1)$ quantum link model (QLM) \cite{Surace2020}. The latter is a quantum link formulation of lattice quantum electrodynamics in $(1+1)$ dimensions, which has recently also been implemented in large-scale ultracold-atom quantum simulators \cite{Yang2020,Zhou2021}. Moreover, scars have recently also been shown to exist in a $\mathbb{Z}_2$ lattice gauge theory (LGT) \cite{Iadecola2020,Sai2022} and in higher-dimensional gauge theories as well \cite{Banerjee2021}. Given that quantum many-body scarring is known not to be stable against perturbations \cite{Surace2021}, and in light of an impressive experimental effort towards realizing gauge theories in synthetic quantum matter devices \cite{Martinez2016,Muschik2017,Klco2018,Barbiero2019,Kokail2019,Barbiero2019,Goerg2019,Schweizer2019,Mil2020,Klco2020,Yang2020,Zhou2021,aidelsburger2021cold}, it is important to investigate methods that may protect scarred dynamics in the presence of errors.

In this work, we explore the potential of recently proposed linear gauge protection schemes \cite{Halimeh2020e,Halimeh2021stabilizing,lang2022disorder} in making scarred dynamics robust against errors in modern quantum-simulation platforms for gauge theories \cite{Pasquans_review,Zohar_review,Dalmonte_review,aidelsburger2021cold,Zohar_NewReview,Bauer_review}. These methods have previously shown reliable stabilization of gauge symmetry up to impressive timescales \cite{Halimeh2020e,vandamme2021reliability,Halimeh2021stabilizing,vandamme2021suppressing}, but have mostly been tested on local observables under generic conditions. It is not clear how they will fare when it comes to protecting fine-tuned features such as quantum many-body scars especially as pertaining the hallmark revivals they incur in the fidelity, which is a global quantity and therefore can be very sensitive to errors. As we show here numerically using time-evolution methods based on Krylov subspaces, linear gauge protection mitigates experimentally prevalent errors and renders scarring robust up to experimentally relevant timescales.

The rest of the paper is organized as follows: In Sec.~\ref{sec:U1QLM}, we show how linear gauge protection can be efficiently used to restore two main types of scarred dynamics in error-prone implementations of the $\mathrm{U}(1)$ QLM. We then show in Sec.~\ref{sec:Z2LGT} the generality of these schemes by demonstrating how they protect scarred dynamics in the $\mathbb{Z}_2$ LGT in the presence of errors inspired from a recent ultracold-atom experiment~\cite{Schweizer2019}. We conclude in Sec.~\ref{sec:conc} and supplement our work in Appendix~\ref{app:supp} with supporting numerical results including those considering different kinds of errors and protection schemes.

\section{$\mathrm{U}(1)$ quantum link model}\label{sec:U1QLM}
\begin{figure}[t!]
    \centering
    \includegraphics[width=\columnwidth]{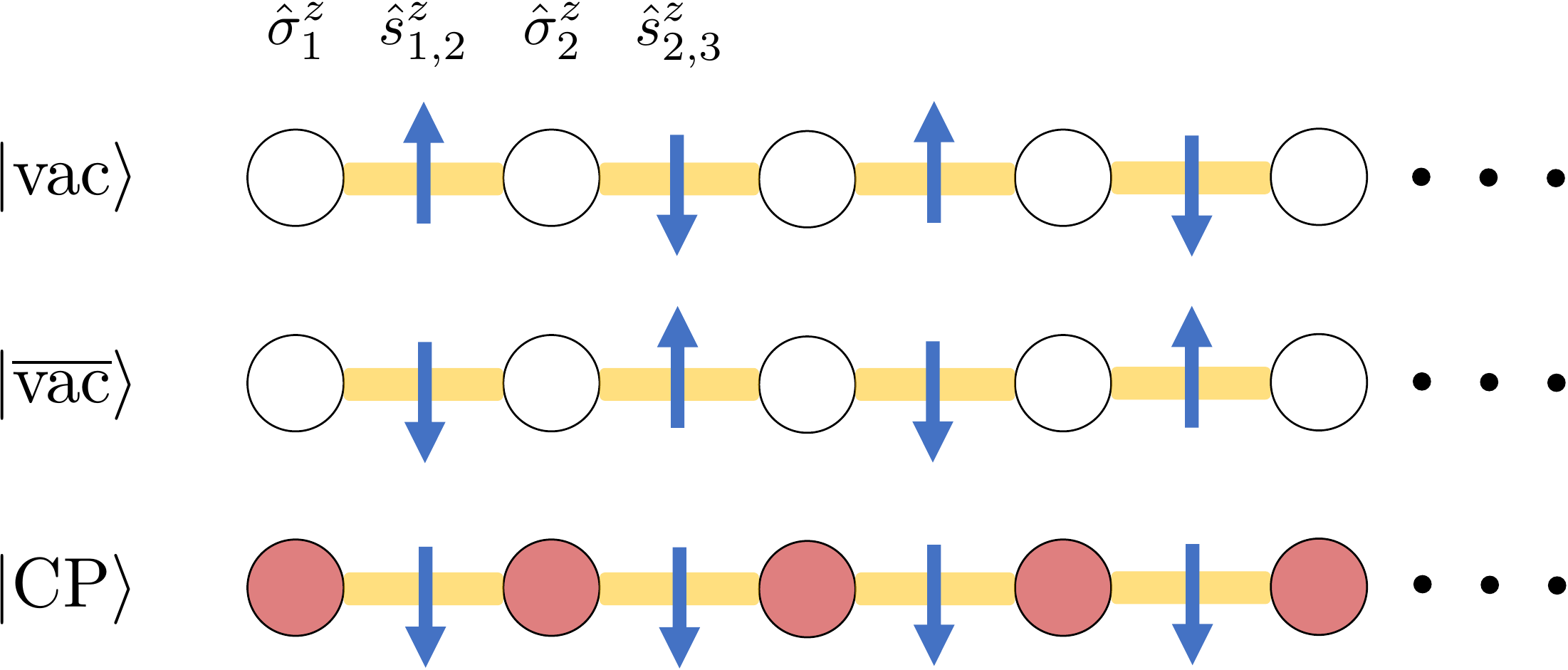}
    \caption{(Color online). Product states relevant to the quenches in the $\mathrm{U}(1)$ quantum link model~\eqref{eq:U1QLM_H0}, the dynamics of which we numerically calculate using Krylov subspace methods. The vacua $\ket{\text{vac}}$ and $\ket{\overline{\text{vac}}}$ are the doubly degenerate ground states of Eq.~\eqref{eq:U1QLM_H0} at $\mu\to\infty$ in the physical sector $\hat{G}_j\ket{\phi}=0,\,\forall j$. They both break the global $\mathbb{Z}_2$ symmetry of this model. They lead to \textit{resonant scarring} when they are quenched with Eq.~\eqref{eq:U1QLM_H0} at $\mu=0$. The charge-proliferated or ``polarized'' state $\ket{\text{CP}}$ is the nondegenerate $\mathbb{Z}_2$-symmetric ground state of Eq.~\eqref{eq:U1QLM_H0} at $\mu\to-\infty$ in the physical sector $\hat{G}_j\ket{\phi}=0,\,\forall j$. It leads to \textit{detuned scarring} when quenched by Hamiltonian~\eqref{eq:U1QLM_H0} with a finite mass around $\mu=-0.84J$. Unless otherwise stated, we employ in our numerics a system size of $L_\text{m}=L=12$ lattice sites and $L_\text{g}=L-1=11$ gauge links with open boundary conditions.}
    \label{fig:U1QLM_InitialStates}
\end{figure}
The spin-$1/2$ $\mathrm{U(1)}$ quantum link model is described by the Hamiltonian \cite{Wiese_review,Yang2016,Kasper2017}
\begin{align}\label{eq:U1QLM_H0}
    \hat{H}_0=J\sum_{j=1}^{L-1}\big(\hat{\sigma}^-_j\hat{s}^+_{j,j+1}\hat{\sigma}^-_{j+1}+\text{H.c.}\big)+\frac{\mu}{2}\sum_{j=1}^L\hat{\sigma}^z_j,
\end{align}
where the first term describes the annihilation and creation of charged matter, whose occupation at matter site $j$ is described by the Pauli matrix $\hat{\sigma}^z_j$, along with the concomitant action from the gauge field, described by the spin-$1/2$ raising operator $\hat{s}^+_{j,j+1}$ at the link between sites $j$ and $j+1$, in order to preserve Gauss's law. The electric field at this link is represented by the spin-$1/2$ operator $\hat{s}^z_{j,j+1}$. The fermionic mass is denoted by $\mu$, and the energy scale is set by $J>0$. The $\mathrm{U}(1)$ QLM is a simplified lattice version of quantum electrodynamics where the $\mathrm{U}(1)$ gauge field is represented by a spin-$1/2$ operator.

\begin{figure*}[t!]
    \centering
    \includegraphics[width=\columnwidth]{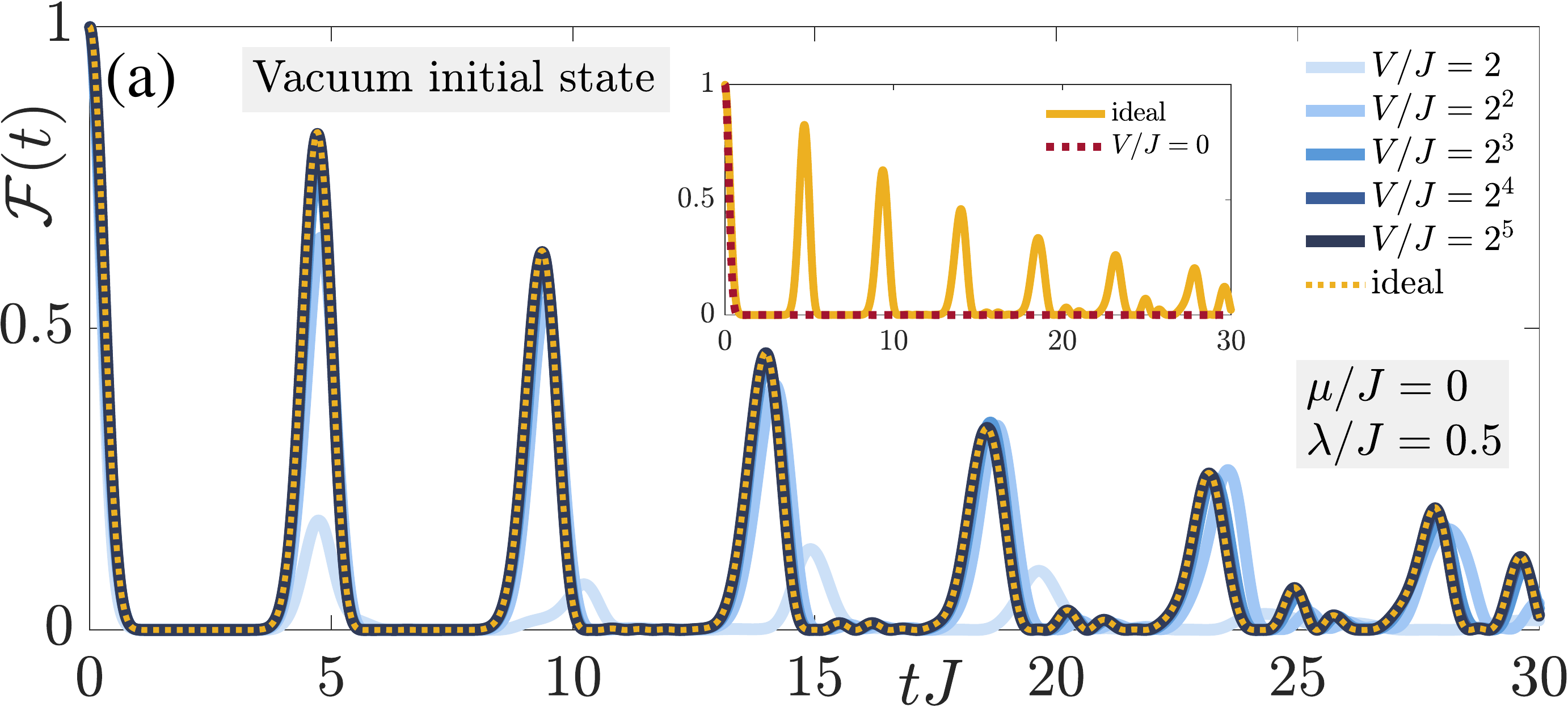}\quad\includegraphics[width=\columnwidth]{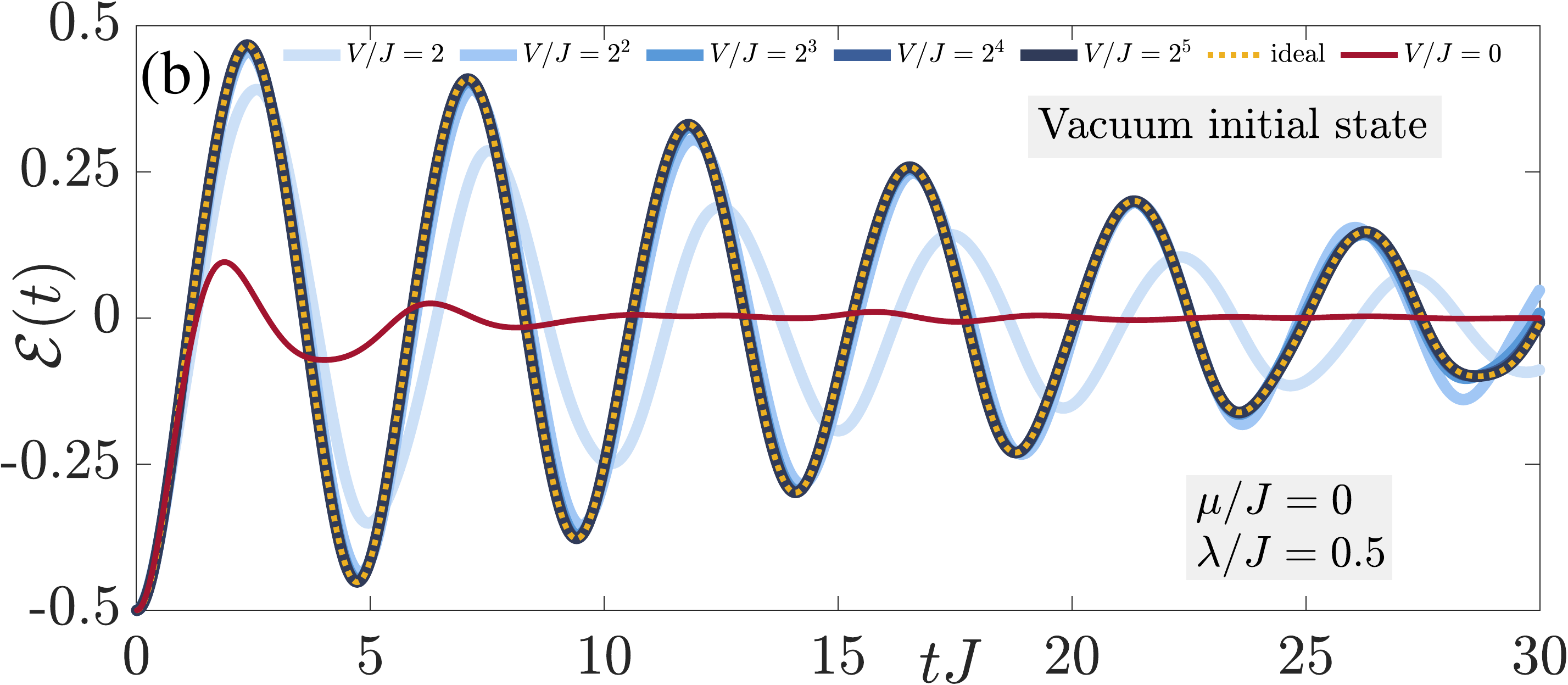}\\
	\vspace{1.1mm}
    \includegraphics[width=\columnwidth]{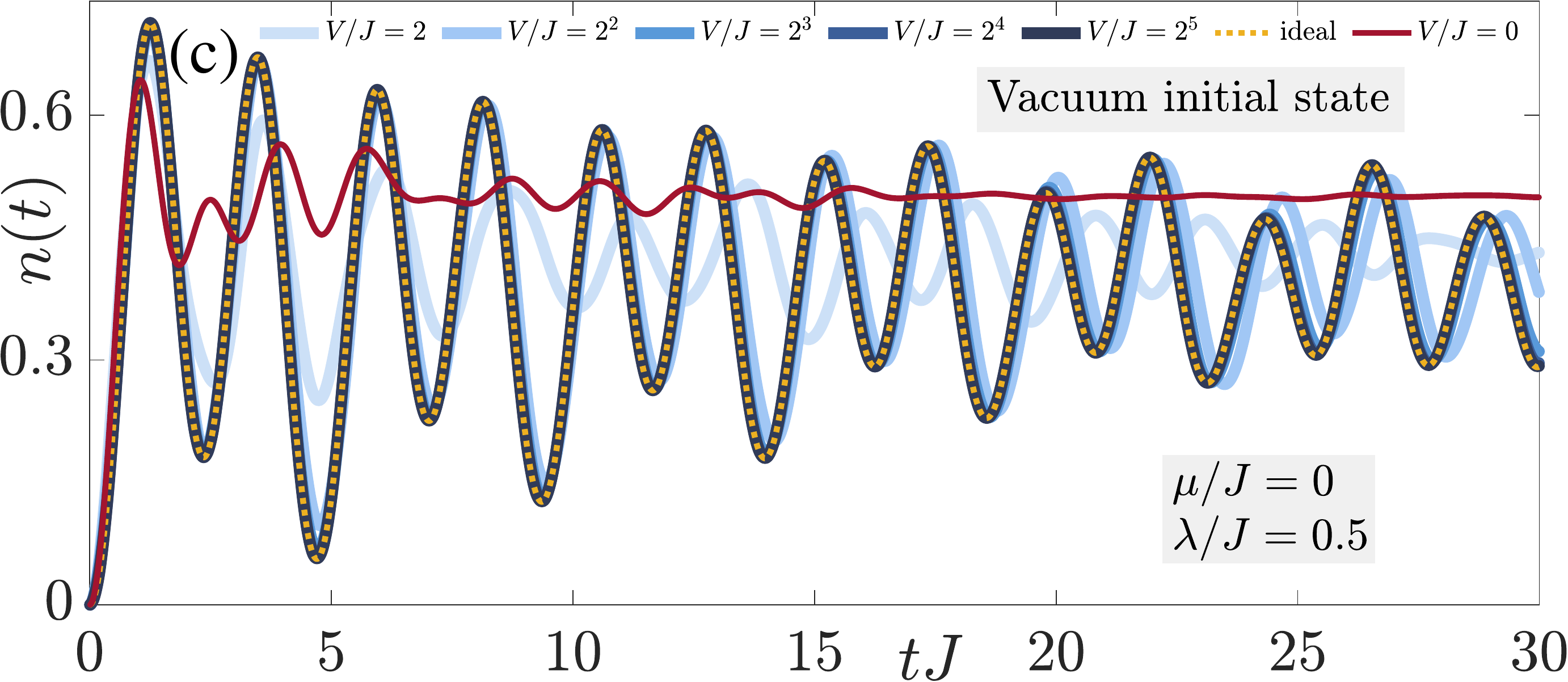}\quad\includegraphics[width=\columnwidth]{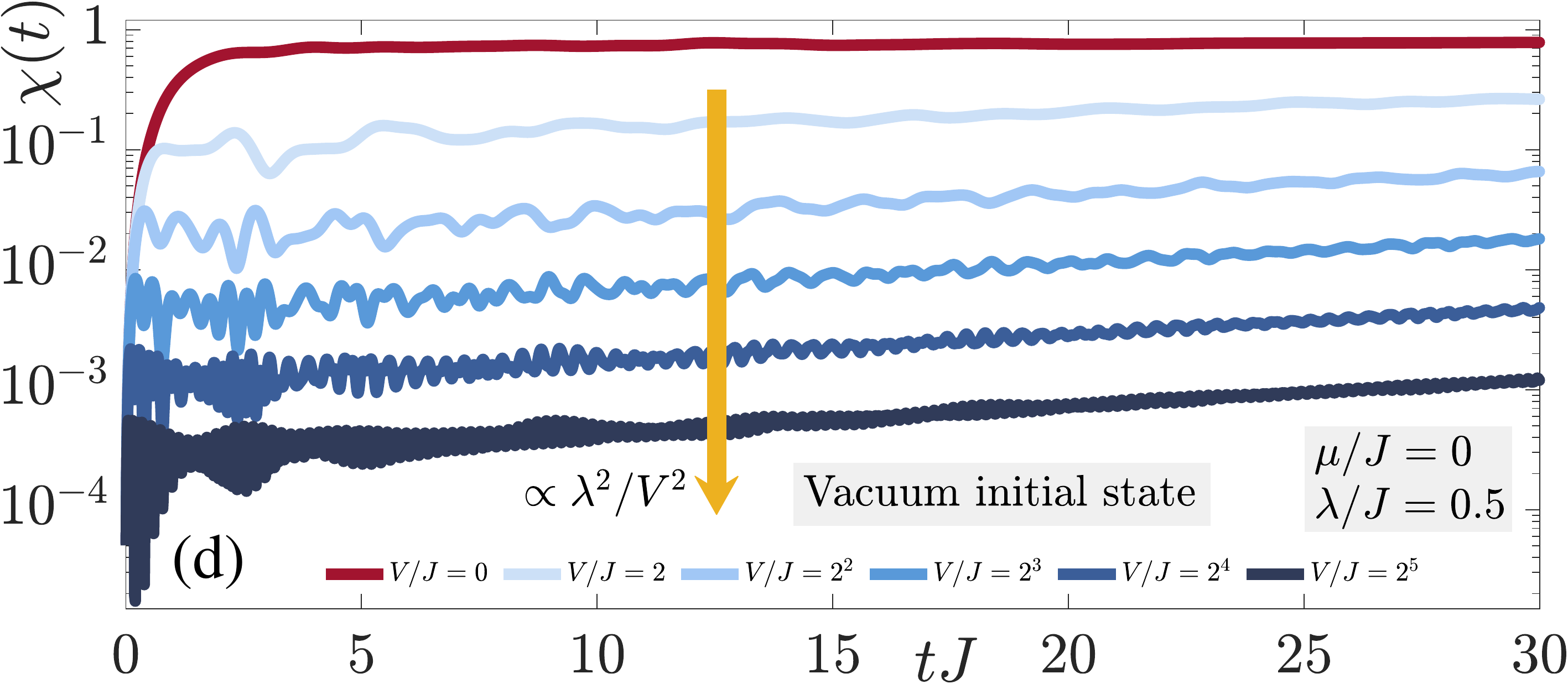}
    \caption{(Color online). Scarring dynamics in the $\mathrm{U}(1)$ quantum link model for a zero-mass quench starting in a vacuum initial state, where unavoidable experimental errors are denoted by $\lambda\hat{H}_1$, against which we protect using the term $V\hat{H}_G=V\sum_{j=2}^{L-1}(-1)^j\hat{G}_j$. The results are obtained for $L_\text{m}=L=12$ matter sites and $L_\text{g}=L-1=11$ gauge links with open boundary conditions using a Krylov-based time-evolution scheme. (a) The fidelity shows periodic revivals in the ideal case (solid yellow curve in the inset, dotted yellow curve in main plot). However, in the presence of errors ($\lambda=0.5J$ here), the fidelity quickly decays (dotted red curve in inset). Upon adding the linear gauge protection term the fidelity is restored to that of the ideal case at sufficiently large protection strength $V$ (different shades of blue). (b,c) Characteristic of many-body scarring is persistent oscillations in local observables such as the (b) electric flux and (c) charge conjugate. Such oscillations are quickly damped in the presence of unprotected errors (red solid curve), but are reliably restored upon employing linear gauge protection (different shades of blue), exactly reproducing the ideal case (yellow dotted curve) at sufficiently large yet experimentally feasible values of $V$ for all investigated evolution times. (d) We can connect the robustness of quantum many-body scarring to the stability of gauge invariance. Whereas the gauge violation quickly grows to a maximal-violation steady state in the case of unprotected errors, upon adding linear gauge protection we see that it gets suppressed $\propto\lambda^2/V^2$ at sufficiently large $V$.}
    \label{fig:u1qlm_vacuum}
\end{figure*}

Gauss's law is described by the discretized generator of the $\mathrm{U}(1)$ gauge symmetry,
\begin{align}\label{eq:U1QLM_Gj}
    \hat{G}_j=(-1)^j\big(\hat{n}_j+\hat{s}^z_{j-1,j}+\hat{s}^z_{j,j+1}\big),
\end{align}
with $1<j<L$ due to open boundary conditions, and where we have used $\hat{n}_j=\big(\hat{\sigma}^z_j+\mathds{1}\big)/2$. The gauge symmetry of Eq.~\eqref{eq:U1QLM_H0} manifests in the latter commuting with Eq.~\eqref{eq:U1QLM_Gj} at every site $j$: $\big[\hat{H}_0,\hat{G}_j\big]=0,\,\forall j$. In the eigenbasis of the generator $\hat{G}_j$, the Hamiltonian $\hat{H}_0$ can be block-diagonalized, with each block corresponding to a unique gauge superselection sector defined by the set of eigenvalues $\mathbf{g}=\big(g_2,g_3,\ldots,g_{L-1}\big)$ of $\hat{G}_j$. The Hamiltonian~\eqref{eq:U1QLM_H0} drives dynamics within each superselection sector, but does not couple different sectors due to its gauge symmetry.

In a realistic implementation of $\hat{H}_0$, unavoidable errors will arise that may explicitly break gauge invariance. For example, experimentally relevant errors for Eq.~\eqref{eq:U1QLM_H0} may take the form \cite{Mil2020}
\begin{align}\label{eq:U1QLM_H1}
    \hat{H}_1=\sum_{j=1}^{L-1}\big(\hat{\sigma}^-_j\hat{\sigma}^-_{j+1}+\hat{\sigma}^+_j\hat{\sigma}^+_{j+1}+\hat{s}^x_{j,j+1}\big),
\end{align}
which represent the creation or annihilation of matter without a change in the electric-field configuration, or vice versa, thereby violating Gauss's law. Such errors pose serious problems to gauge-theory simulations in general, and controlling them becomes crucial.

Recently, a gauge symmetry-protection scheme linear in the gauge-symmetry generators has been proposed, $V\hat{H}_G=\sum_{j=2}^{L-1}c_j\hat{G}_j$, where the coefficients $c_j$ can be tailored such that the $\mathrm{U}(1)$ gauge symmetry generated by $\hat{G}_j$ is stabilized in the faulty theory $\hat{H}=\hat{H}_0+\lambda\hat{H}_1+V\hat{H}_G$ up to well-defined timescales \cite{Halimeh2020e}. The protection sequence $c_j$ is said to be compliant if it is integer and satisfies $\sum_jc_j\big(g_j-g_j^\text{tar}\big)=0\iff g_j=g_j^\text{tar},\,\forall j\in\{2,\ldots,L-1\}$, where $\mathbf{g}^\text{tar}=\big(g_2^\text{tar},g_3^\text{tar},\ldots,g_{L-1}^\text{tar}\big)$ is the target gauge sector. At sufficiently large $V$, the linear protection term stabilizes gauge invariance up to times exponential in $V$. Since we are interested in large-scale implementations of the $\mathrm{U}(1)$ QLM such as those recently realized in ultracold atoms \cite{Yang2020,Zhou2021}, such a compliant sequence is not ideal because it grows exponentially with system size for a fixed $V$. In keeping with experimental relevance, the \textit{noncompliant} sequence $c_j=(-1)^j$ has been employed in numerical benchmarks, showing excellent stabilization of gauge invariance up to all accessible evolution times in both exact diagonalization and infinite matrix product state calculations \cite{Halimeh2020e,vandamme2021reliability}.

Let us now investigate the efficacy of the single-body staggered gauge protection scheme, \begin{align}\label{eq:HG}
V\hat{H}_G=V\sum_{j=2}^{L-1}(-1)^j\hat{G}_j,
\end{align}
in stabilizing quantum many-body scarring in the $\mathrm{U}(1)$ QLM. Even though this scheme has been shown to protect the dynamics of local observables \cite{Halimeh2020e,vandamme2021reliability}, how it will fare when it comes to global quantities such as the fidelity is still an open question. Quantum many-body scars require specific fine-tuning in the initial state and the quench Hamiltonian in order to emerge in an experiment, and the fidelity has been a main quantity used to probe their existence \cite{Su2022}. For these reasons, it is important to assess how well modern protection schemes can stabilize them.

\subsection{Resonant scarring}
We now prepare the system in the vacuum initial state $\ket{\text{vac}}$ depicted in Fig.~\ref{fig:U1QLM_InitialStates}. This is one of two doubly degenerate ground states of Eq.~\eqref{eq:U1QLM_H0} at $\mu\to\infty$ in the physical sector $\hat{G}_j\ket{\phi}=0,\,\forall j$. Quenching this initial state with the Hamiltonian~\eqref{eq:U1QLM_H0} at $\mu=0$ (resonantly) is known to lead to scarring behavior, and was first observed in the form of persistent oscillations in local observables in a Rydberg-atom setup~\cite{Bernien2017,Surace2020}. We shall refer to this as \textit{resonant scarring}. Let us now numerically investigate how scarring dynamics is affected in the presence of errors such as~\eqref{eq:U1QLM_H1}, and how the single-body protection scheme~\eqref{eq:HG} can mitigate them. Unless otherwise specified, our results are obtained from time-evolution methods based on Krylov subspaces \cite{Moler2003,EXPOKIT}, for a chain of $L_\text{m}=L=12$ matter sites and $L_\text{g}=L-1=11$ gauge links, with open boundary conditions. This is the system size at which finite-size effects are no longer relevant over the times we probe \cite{Turner2018}.

We first compute the dynamics of the fidelity,
\begin{align}\label{eq:fidelity}
    \mathcal{F}(t)=\big\lvert\braket{\text{vac}}{\psi(t)}\big\rvert^2,
\end{align}
where $\ket{\psi(t)}=e^{-i\hat{H}t}\ket{\text{vac}}$ and $\hat{H}=\hat{H}_0+\lambda\hat{H}_1+V\sum_j(-1)^j\hat{G}_j$,
in Fig.~\ref{fig:u1qlm_vacuum}(a). In the idealized case of no errors, we see consistent revivals up to all accessible evolution times indicating nonthermal behavior; see solid yellow curve in the inset, dotted yellow curve in main plot. In between two revivals, there is ``state transfer'' from the initial state $\ket{\text{vac}}$ to the second vacuum $\ket{\overline{\text{vac}}}$ (see Fig.~\ref{fig:U1QLM_InitialStates}), and back to $\ket{\text{vac}}$. The revival period is determined by the energy spacing between the roughly equally spaced scar eigenstates of the quench Hamiltonian~\cite{Serbyn2020}. Upon introducing even perturbative errors (results shown for $\lambda=0.5J$) without protection ($V=0$), the fidelity exhibits ergodic behavior, quickly decaying to zero; see dotted red curve in inset. Upon employing the gauge protection scheme~\eqref{eq:HG}, however, we find that nonthermal behavior is restored in the fidelity (different shades of blue), and even quantitative agreement with the ideal case over all evolution times is achieved at moderate values of the protection strength $V\gtrsim 8J$.

A prominent feature of scarred dynamics is persistent oscillations in local observables \cite{Bernien2017}. As such, we calculate the dynamics of the electric flux and the chiral condensate
\begin{subequations}
\begin{align}\label{eq:flux}
    \mathcal{E}(t)&=\frac{1}{L-1}\sum_{j=1}^{L-1}(-1)^j\bra{\psi(t)}\hat{s}^z_{j,j+1}\ket{\psi(t)},\\\label{eq:CC}
    n(t)&=\frac{1}{L}\sum_{j=1}^{L}\bra{\psi(t)}\hat{n}_j\ket{\psi(t)},
\end{align}
\end{subequations}
respectively. The electric flux serves as an order parameter, which is nonzero when the global $\mathbb{Z}_2$ symmetry of Eq.~\eqref{eq:U1QLM_H0} is spontaneously broken. The chiral condensate measures how strongly the chiral symmetry related to fermions in the model is spontaneously broken. The corresponding results are shown in Fig.~\ref{fig:u1qlm_vacuum}(b,c). We see in the case of unprotected errors ($V=0$) that both observables show quick relaxation to their thermal value (red solid curves). Upon adding the gauge protection term~\eqref{eq:HG}, we find that oscillations are revived (different shades of blue), with excellent quantitative agreement with the ideal case (yellow dotted curves) already at $V\approx8J$ up to all investigated evolution times, indicating robust scarring dynamics.

It is interesting to note in the case of robust scarring ($V\gtrsim 8J$) that the period of oscillation for the electric flux~\eqref{eq:flux} is roughly the same as that of the fidelity~\eqref{eq:fidelity}, but the chiral condensate~\eqref{eq:CC} has half the period of the latter. The fidelity revivals indicate that the wave function is returning very close to its initial state $\ket{\text{vac}}$. However, in-between two revivals, the wave function very closely approaches the second degenerate vacuum $\ket{\overline{\text{vac}}}$, which corresponds to an infinitesimally small fidelity in the middle of two consecutive revivals. Since the electric flux is the order parameter associated with the spontaneous breaking of a global $\mathbb{Z}_2$ symmetry in this model, it is at a local maximum when the wave function approaches the second vacuum $\ket{\overline{\text{vac}}}$, itself of maximal order (see Fig.~\ref{fig:U1QLM_InitialStates}). When the fidelity revives, however, the order parameter is again at a local minimum, corresponding to the vacuum $\ket{\text{vac}}$, which has minimal order and in which the system was initially prepared. On the other hand, the chiral condensate is not an order parameter, and therefore need not have the same period as the fidelity. Instead, we find that the chiral condensate has half the period of the latter. This makes sense when noticing that the matter-field configuration must be the same for both vacua $\ket{\text{vac}}$ and $\ket{\overline{\text{vac}}}$ (zero chiral condensate), as shown in Fig.~\ref{fig:U1QLM_InitialStates}. As such, when the wave function is near the second vacuum, it must give rise to a local minimum in the chiral condensate, just like at the initial state, but the chiral condensate must also be at a local minimum when it goes back to the first vacuum. This therefore requires the chiral condensate to have double the frequency of the fidelity and electric flux.

We now investigate the connection between the robustness of scarring and the stability of gauge invariance in the dynamics of this quench. For that purpose, we compute the gauge violation
\begin{align}~\label{eq:viol}
    \chi(t)=\frac{1}{L-2}\sum_{j=2}^{L-1}\bra{\psi(t)}\big(\hat{G}_j-g_j^\text{tar}\big)^2\ket{\psi(t)},
\end{align}
the dynamics of which are shown in Fig.~\ref{fig:u1qlm_vacuum}(d). We recall here that we work in the gauge sector $g_j^\text{tar}=0,\,\forall j$. In all cases, one can show through time-dependent perturbation theory \cite{Halimeh2020a} that the gauge violation~\eqref{eq:viol} grows $\propto\lambda^2t^2$ at early times in the presence of errors, $\lambda=0.5J$. Without any protection, the gauge violation grows until it plateaus into a maximal-violation steady state at a timescale $\propto1/\lambda$. Once the gauge protection is turned on, however, we see that beginning at a timescale $\propto1/V$ there is a suppression of the value of this plateau $\propto\lambda^2/V^2$, as can be derived in degenerate perturbation theory \cite{Halimeh2020a}. The behavior of the gauge-violation dynamics shows that the robustness of scarring is directly connected to a stable gauge symmetry.

\subsection{Detuned scarring}
In a recent ultracold-atom experiment \cite{Su2022}, a tilted Bose--Hubbard optical lattice that maps onto the $\mathrm{U(1)}$ QLM in the physical sector $g_j^\text{tar}=0,\,\forall j$, has been employed to investigate scarring behavior for a detuned quench starting in a Mott-insulator state. We shall refer to this as \textit{detuned scarring}. This is equivalent to quenching the charge-proliferated state $\ket{\text{CP}}$ in Fig.~\ref{fig:U1QLM_InitialStates} with Hamiltonian~\eqref{eq:U1QLM_H0} at nonzero $\mu$.  The state $\ket{\text{CP}}$ is the ground state of Eq.~\eqref{eq:U1QLM_H0} at $\mu\to-\infty$ in the physical sector. 

\begin{figure}[t!]
    \centering
    \includegraphics[width=\columnwidth]{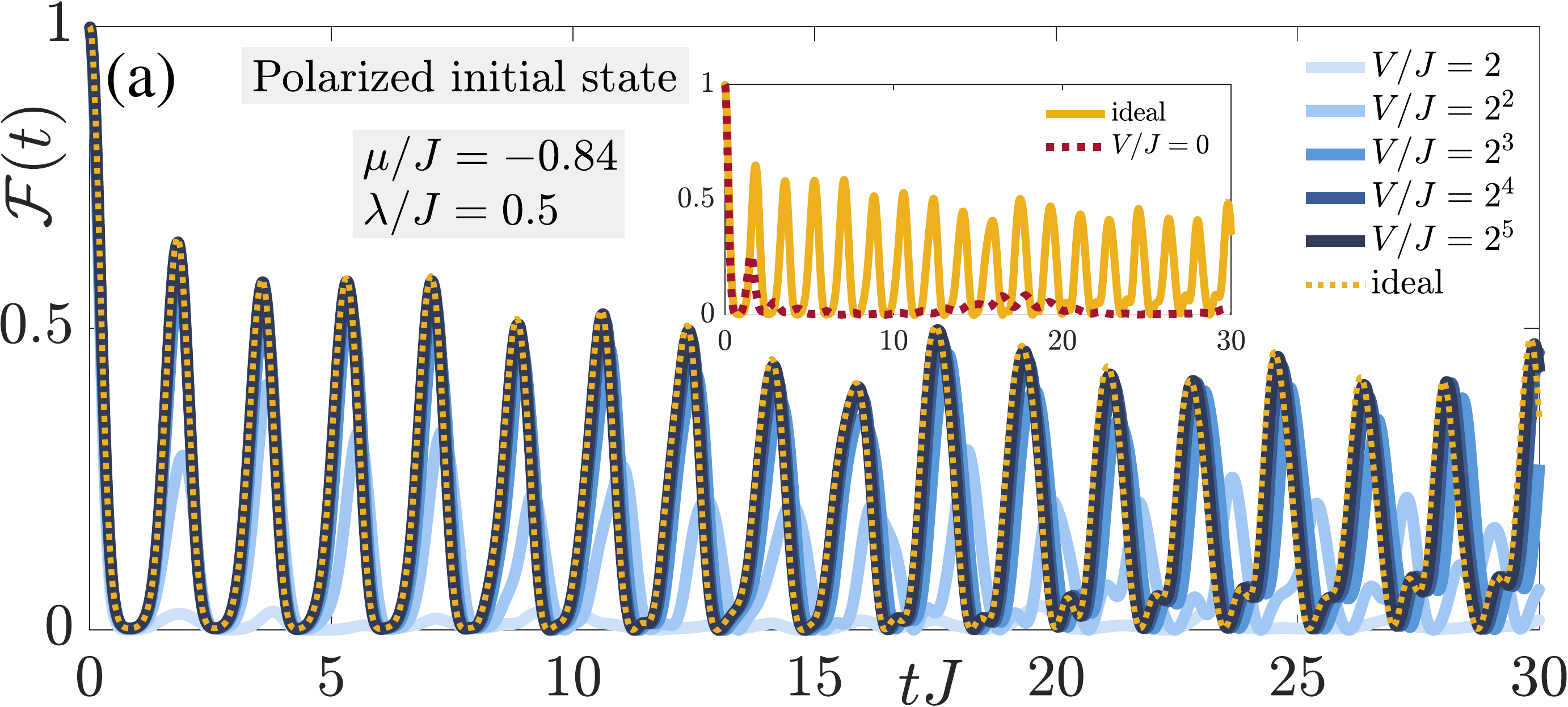}\\
	\vspace{1.1mm}
    \includegraphics[width=\columnwidth]{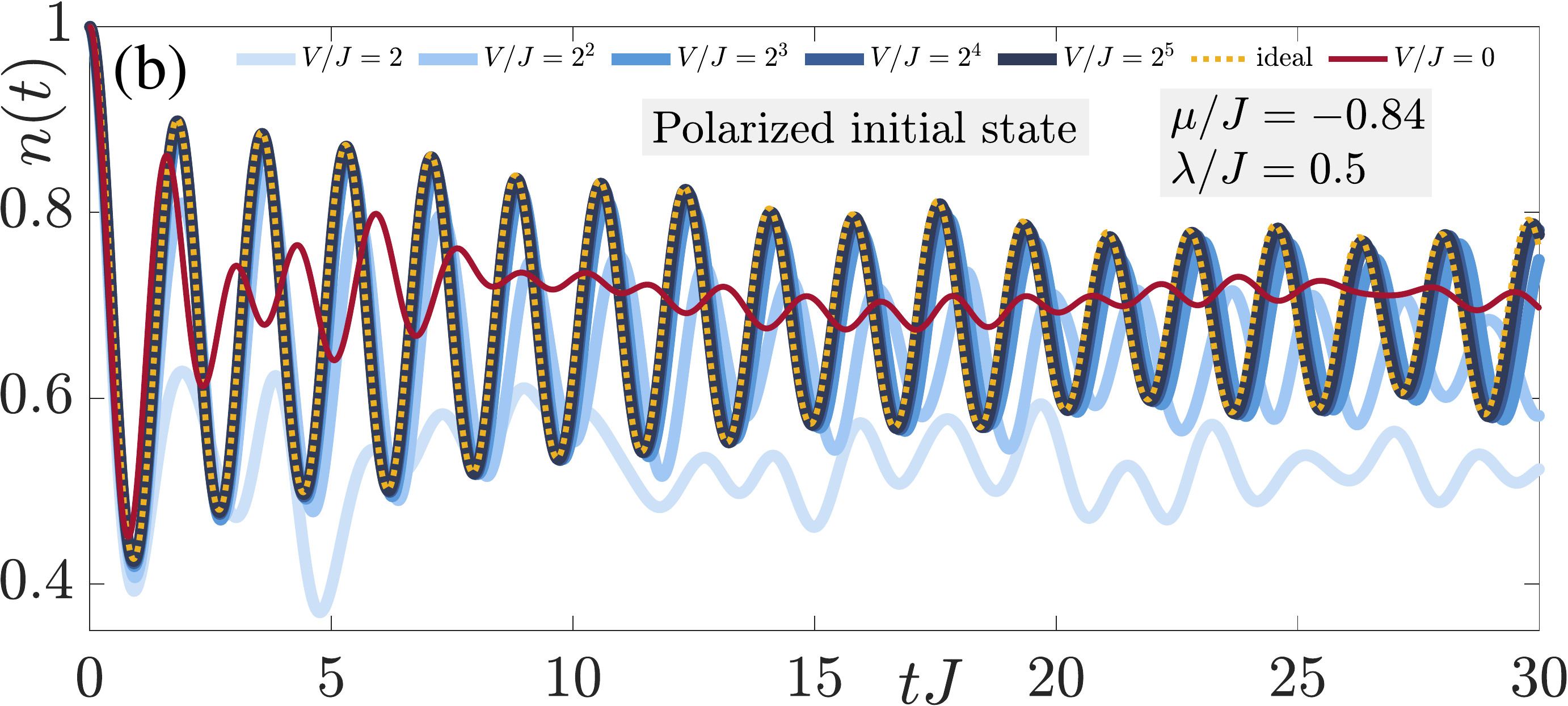}\\
	\vspace{1.1mm}
    \includegraphics[width=\columnwidth]{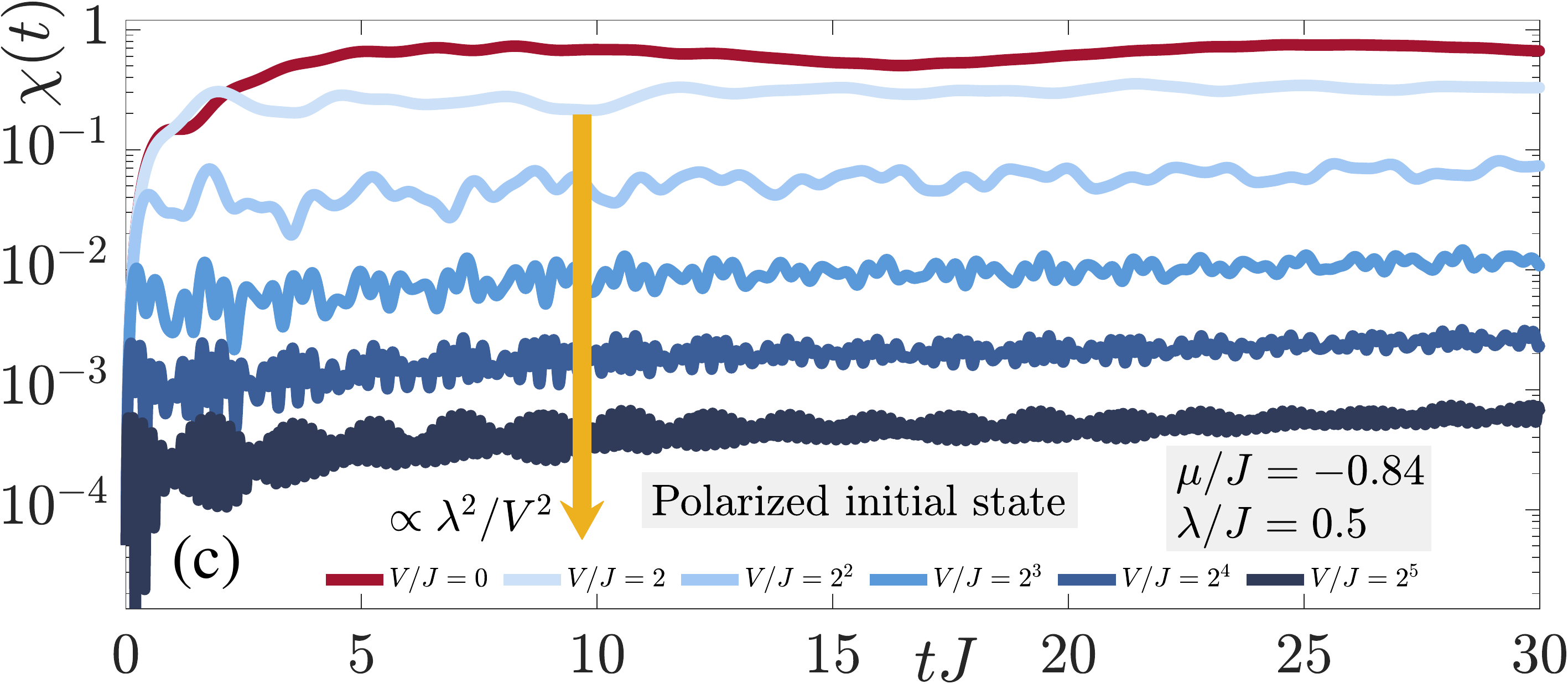}
    \caption{(Color online). Same as Fig.~\ref{fig:u1qlm_vacuum} but for a nonzero-mass quench starting in the charge-proliferated (polarized) initial state. The qualitative picture remains unchanged.}
    \label{fig:u1qlm_polarized}
\end{figure}

We perform this quench in our numerical simulations, where we start in the charge-proliferated state $\ket{\text{CP}}$ and quench it with Hamiltonian~\eqref{eq:U1QLM_H0} at $\mu=-0.84J$, which is within the optimal range to generate scarred dynamics for this initial state \cite{Su2022}. We see consistent revivals in the fidelity dynamics for the ideal case in Fig.~\ref{fig:u1qlm_polarized}(a)---yellow solid curve in inset, yellow dotted curve in main plot---indicating nonthermal scarred behavior. However, as soon as errors are included (here we set $\lambda=0.5J$), the fidelity quickly decays and remains mostly vanishing throughout the whole time evolution (red dotted curve in inset). As in the case of its resonant counterpart, we can make detuned scarring also robust by introducing the gauge protection scheme in Eq.~\eqref{eq:HG}, the corresponding dynamics of which are shown in different shades of blue. Even at the smallest considered value $V=2J$ of the protection strength, the fidelity displays small revivals up to relatively long evolution times. Already at the experimentally friendly value of $V=16J$, we find very good quantitative agreement with the ideal case (yellow dotted curve).

Turning to local observables, we see the same qualitative picture in the dynamics of the chiral condensate~\eqref{eq:CC} in Fig.~\ref{fig:u1qlm_polarized}(b). In the ideal case, it shows persistent oscillations over all accessible evolution times, in keeping with scarred dynamics (yellow dotted curve). In the presence of errors without protection, the dynamics shows faster relaxation, with only strongly damped oscillations (red solid curve). Upon employing gauge protection, however, the revivals in the chiral condensate become prominent again. For the moderate value of $V=16J$, the quantitative agreement with the ideal case is very good over the accessible evolution times. Unlike in the case of resonant scarring, here we find that the chiral condensate has the same period as the fidelity. This lies in the fact that the charge-proliferated state is nondegenerate. This means that half-way between two revivals in the fidelity, the wave function is not required to exhibit the same matter-field configuration by passing through a second degenerate state.

Just as in the case of resonant scarring, we find that the stability of gauge symmetry is directly connected to the robustness of quantum many-body scars in this case as well. As shown in Fig.~\ref{fig:u1qlm_polarized}(c), the gauge violation is suppressed into a plateau of value $\propto\lambda^2/V^2$ at sufficiently large $V$, and this reflects in the robustness of the corresponding scarring dynamics in the fidelity and chiral condensate in Fig.~\ref{fig:u1qlm_polarized}(a,b).

\subsection{Effective Zeno Hamiltonian}
The concept of quantum Zeno dynamics, usually associated with nonunitary dynamics involving frequent measurements on the system, can also be employed in the context of \textit{strong continuous coupling}, which is unitary and involves adding a term to the Hamiltonian with a dominant energy scale \cite{facchi2002quantum,facchi2004unification,facchi2009quantum,burgarth2019generalized}. It can then be employed to rigorously show that starting in the physical sector, the quench dynamics under the faulty theory $\hat{H}=\hat{H}_0+\lambda\hat{H}_1+V\hat{H}_G$ is effectively reproduced by the Zeno Hamiltonian \cite{Halimeh2020e}
\begin{align}\label{eq:QZE}
    \hat{H}_Z=\hat{H}_0+\lambda\hat{\mathcal{P}}_0\hat{H}_1\hat{\mathcal{P}}_0,
\end{align}
where $\hat{\mathcal{P}}_0$ is the projector onto the physical sector, up to a timescale $\propto V/(LV_0)^2$ in a worst-case scenario, with $V_0$ an energy term comprised of a linear sum in $J$, $\mu$, and $\lambda$ \cite{Halimeh2020e}. Here, the eigenprojectors of $V\hat{H}_G$ effectively split the Hilbert space into different nonintersecting quantum Zeno subspaces, between which couplings are suppressed up to times at least linear in $V$ \cite{Halimeh2020e}.

For the particular experimentally relevant \cite{Mil2020} error in Eq.~\eqref{eq:U1QLM_H1}, $\hat{\mathcal{P}}_0\hat{H}_1\hat{\mathcal{P}}_0=0$, and therefore the Zeno Hamiltonian is exactly the ideal theory: $\hat{H}_Z=\hat{H}_0$. This allows the excellent quantitative, in addition to the qualitative, agreement at sufficiently large $V$ between the scarring dynamics under the faulty theory and that under the ideal theory over the accessible evolution times. However, we emphasize here that even if $\hat{\mathcal{P}}_0\hat{H}_1\hat{\mathcal{P}}_0\neq0$, our linear gauge protection scheme will still suppress gauge violations all the same, except that the effective Zeno Hamiltonian will then be a renormalized version of $\hat{H}_0$.

It is to be noted here that while the concept of quantum Zeno dynamics is known to stabilize gauge theories \cite{Halimeh2020e}, it has hitherto been an open question as to whether also highly fine-tuned phenomena such as quantum many-body scars can be protected under this scheme. Our results have resolved this question, showing that experimentally feasible single-body gauge terms such as those in Eq.~\eqref{eq:HG} can make scars robust up to all experimentally relevant timescales. Indeed, the evolution times we access in our numerical simulations are similar to or even longer than those achieved in modern synthetic quantum matter setups observing scars \cite{Lukin2019,Su2022}.

\section{$\mathbb{Z}_2$ lattice gauge theory}\label{sec:Z2LGT}
Let us now turn our attention to another model that has received a lot of recent theoretical \cite{Zohar2017,Barbiero2019,Borla2020,Yang2020fragmentation,kebric2021confinement,Borla2021} and experimental \cite{Goerg2019,Schweizer2019} attention, the $\mathbb{Z}_2$ lattice gauge theory (LGT). In $(1+1)$ dimensions, it is described by the Hamiltonian
\begin{equation}
\hat{H}_0=\sum_{j=1}^{L-1}\big[J\big(\hat{b}_j^\dagger\hat{\sigma}_{j,j+1}^z\hat{b}_{j+1}+\text{H.c.}\big)-h\hat{\sigma}_{j,j+1}^x\big],
\label{h_Z2}
\end{equation}
where $\hat{b}_j^\dagger,\hat{b}_j$ are the creation and annihilation operators of hard-core bosons on site $j$, respectively, with boson-number operator $\hat{n}_j=\hat{b}_j^\dagger\hat{b}_j$,  $J$ describes the hopping processes mediated by the gauge field $\hat{\sigma}^z_{j,j+1}$, a Pauli matrix, which resides on the link connecting the two nearest-neighbor sites $j$ and $j+1$, and $h$ is the strength of an effective electric field $\hat{\sigma}_{j,j+1}^x$, also a Pauli matrix, on this link, imparting dynamics on the gauge field. Here, we further fix the bosonic density at half-filling $\bar{n}=N/L=1/2$, with $N=L/2$ the number of bosons and $L$ the number of lattice sites. The Hamiltonian~\eqref{h_Z2} commutes with the local 
generators of the $\mathbb{Z}_2$ gauge symmetry,
\begin{align}\label{eq:Z2LGT_Gj}
\hat{G}_j=\hat{\sigma}_{j-1,j}^x(-1)^{\hat{n}_j}\hat{\sigma}_{j,j+1}^x.
\end{align}
Gauge-invariant states satisfy $\hat{G}_j\ket{\phi}=\pm\ket{\phi}$, i.e., there are locally only two possible background charges $g_j=\pm1$ corresponding to the underlying $\mathbb{Z}_2$ gauge symmetry. 

Equation~\eqref{h_Z2} can be rewritten in terms of gauge-invariant local spin operators as \cite{Borla2020}
\begin{align}\nonumber
H=\sum_{j=1}^{L-1}\bigg[&\frac{J}{2}\big(\hat{Z}_{j,j+1}-\hat{X}_{j-1,j}\hat{Z}_{j,j+1}\hat{X}_{j+1,j+2}\big)\\\label{H_spin}
&-h\hat{X}_{j,j+1}\bigg],
\end{align}
where
\begin{subequations}
\begin{align}
&\hat{X}_{j,j+1}=\hat{\sigma}_{j,j+1}^x,\\
&\hat{Z}_{j,j+1}=\big(\hat{b}_j^\dagger-\hat{b}_j\big)\hat{\sigma}^z_{j,j+1}\big(\hat{b}_{j+1}^\dagger+\hat{b}_{j+1}\big).
\end{align}
\end{subequations}
Notice that the relation $\hat{n}_j=\big(1-\hat{X}_{j-1,j}\hat{X}_{j,j+1}\big)/2$ holds, and therefore the number of bosons in Hamiltonian~\eqref{h_Z2} is equivalent to the number of domain walls in the Hamiltonian~\eqref{H_spin}, which we keep as a conserved quantum number corresponding to a global $\mathrm{U}(1)$ symmetry.

\begin{figure}[t!]
    \centering
    \includegraphics[width=\columnwidth]{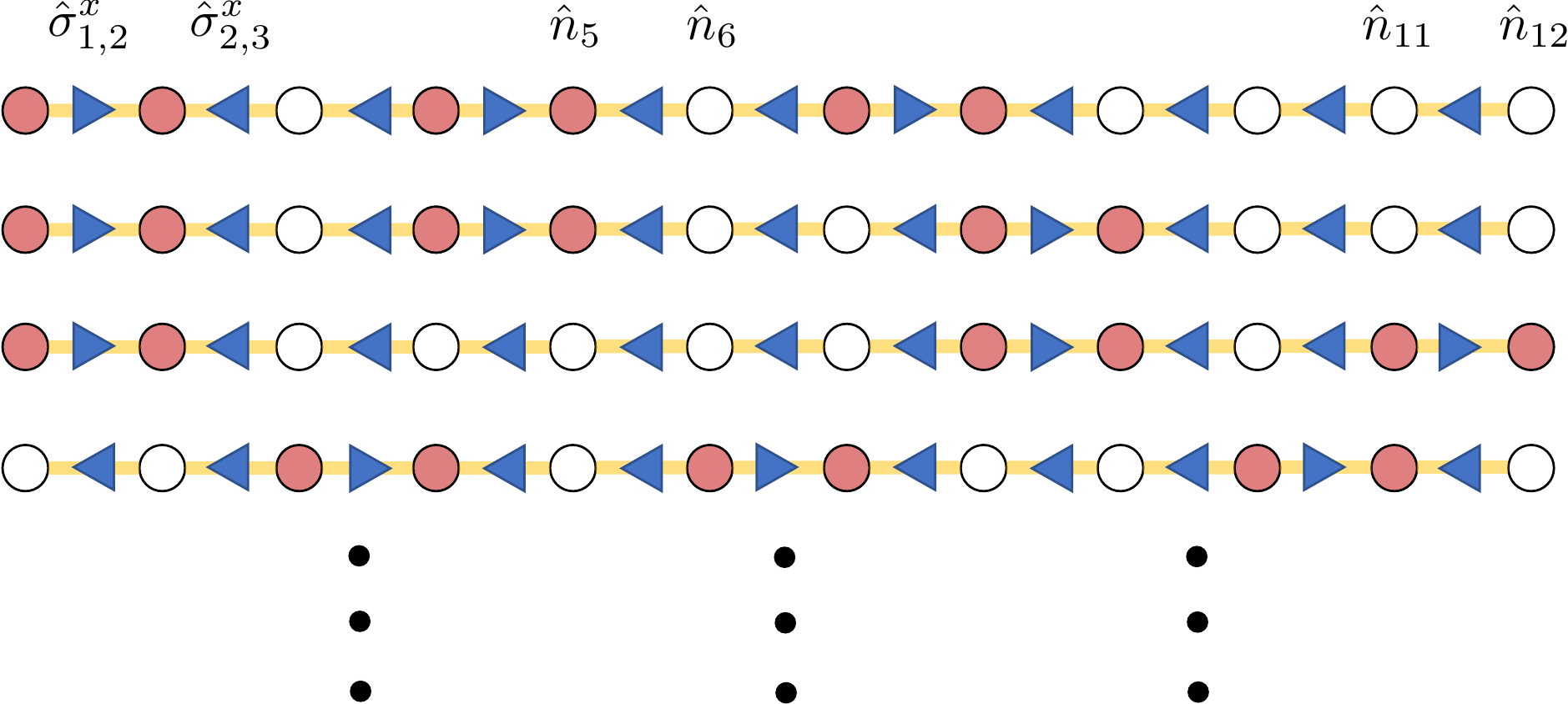}
    \caption{(Color online). Sample product states ($4$ out of $35$ for the system size we use, $L_\mathrm{m}=L=12$ lattice sites and $L_\mathrm{g}=L-1=11$ gauge links with open boundary conditions), an equal superposition of which gives an initial state that will lead to scarred dynamics when quenched by the ideal $\mathbb{Z}_2$ LGT Hamiltonian~\eqref{h_Z2}. All these product states live in the target gauge sector $g_j^\text{tar}=+1,\,\forall j$. Each contains three pairs of adjacent bosons (half-filling) where any two pairs of bosons are separated by at least one empty lattice site.}
    \label{fig:Z2LGT_InitialState}
\end{figure}

Upon defining the model on a dual lattice where the links are represented by sites, and upon a spin-basis rotation ($z\leftrightarrow x$), the Hamiltonian in Eq.~\eqref{H_spin} becomes exactly the same model considered in Ref.~\cite{Iadecola2020}. Here, two towers of quantum many-body scars have been derived,
\begin{equation}
\ket{S_m^k}=\frac{1}{m!\sqrt{{\cal{N}}(L,m)}}\Big[\big(\hat{Q}^k\big)^\dagger\Big]^m\ket{\Omega^k},
\label{qmbs}
\end{equation}
with $k=1,2$, $\ket{\Omega^1}=\ket{\downarrow,\downarrow,\ldots,\downarrow}$, $\ket{\Omega^2}=\ket{\uparrow,\uparrow,\ldots,\uparrow}$, and $\hat{Z}_{j,j+1}\ket{\Omega^k}=(-1)^k\ket{\Omega^k}$. The parameter $m$ refers to the number of magnons, defined as the number of spins-$\uparrow$ in $\ket{\Omega^k}$, and it automatically fixes the number of domain walls (or bosons) through the relation $m=2N$. In the case of open boundary conditions as we adopt in this work, ${\cal{N}}(L,m)=\binom{L-m-1}{m}$,  $(\hat{Q}^k)^\dagger=\sum_j\hat{P}^k_{j-1}\hat{\sigma}_j^{\alpha_k}\hat{P}^k_{j+1}$, where $j$ runs from the second to the second last link, and $\alpha_{1(2)}=\pm$, $\hat{\sigma}^{\pm}_j=\big(\hat{\sigma}^x_j\pm i \hat{\sigma}^y_j\big)$ with $\hat{P}^k_{j}=\big[1+(-1)^k\hat{Z}_{j}\big]$. Upon imposing Gauss's law $\hat{G}_j\ket{\phi}=g_j^\text{tar}\ket{\phi}$, the states~\eqref{qmbs} automatically become quantum many-body scars for the Hamiltonian~\eqref{h_Z2}. By looking at the structure of such a state in addition to the constraints imposed by Gauss's law, Eq.~\eqref{qmbs} describes quantum states where pairs of bosons sitting in nearest-neighbor sites can never reach an interparticle distance smaller than two lattice sites; see Fig.~\ref{fig:Z2LGT_InitialState}, where we show a few samples product states. In total there will be $35$ such product states for the system size we use in this paper ($L=12$ sites and $11$ gauge links) in the target sector $g_j^\text{tar}=+1$.

\begin{figure}[t!]
    \centering
    \includegraphics[width=\columnwidth]{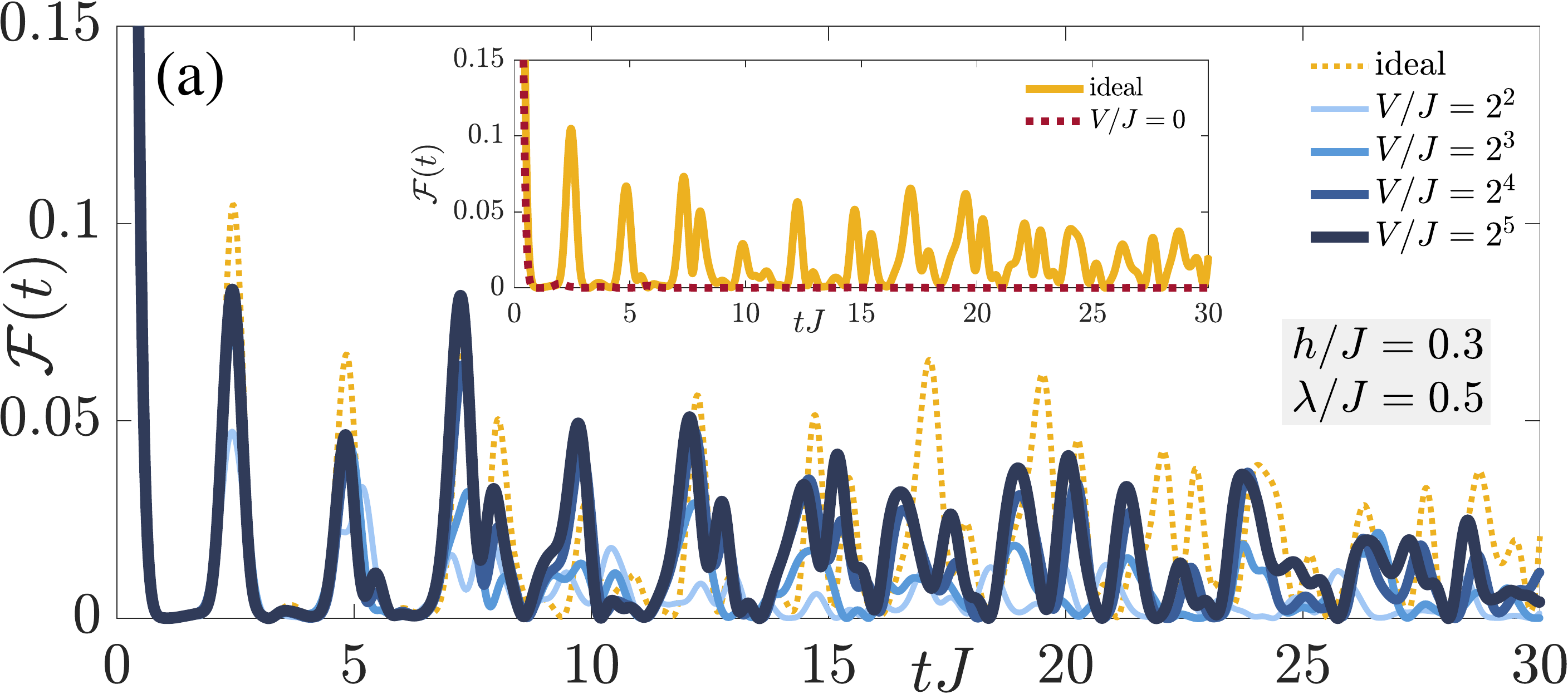}\\
	\vspace{1.1mm}
    \includegraphics[width=\columnwidth]{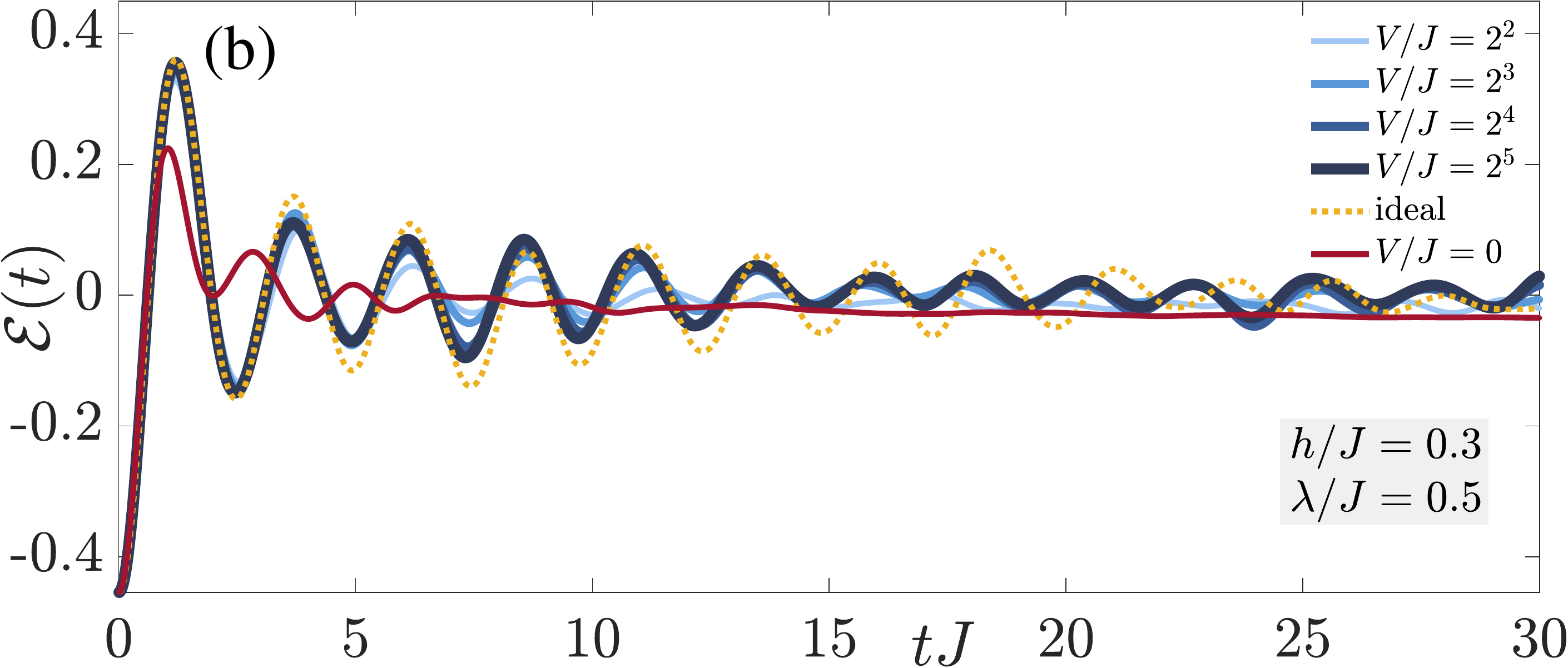}\\
	\vspace{1.1mm}
    \includegraphics[width=\columnwidth]{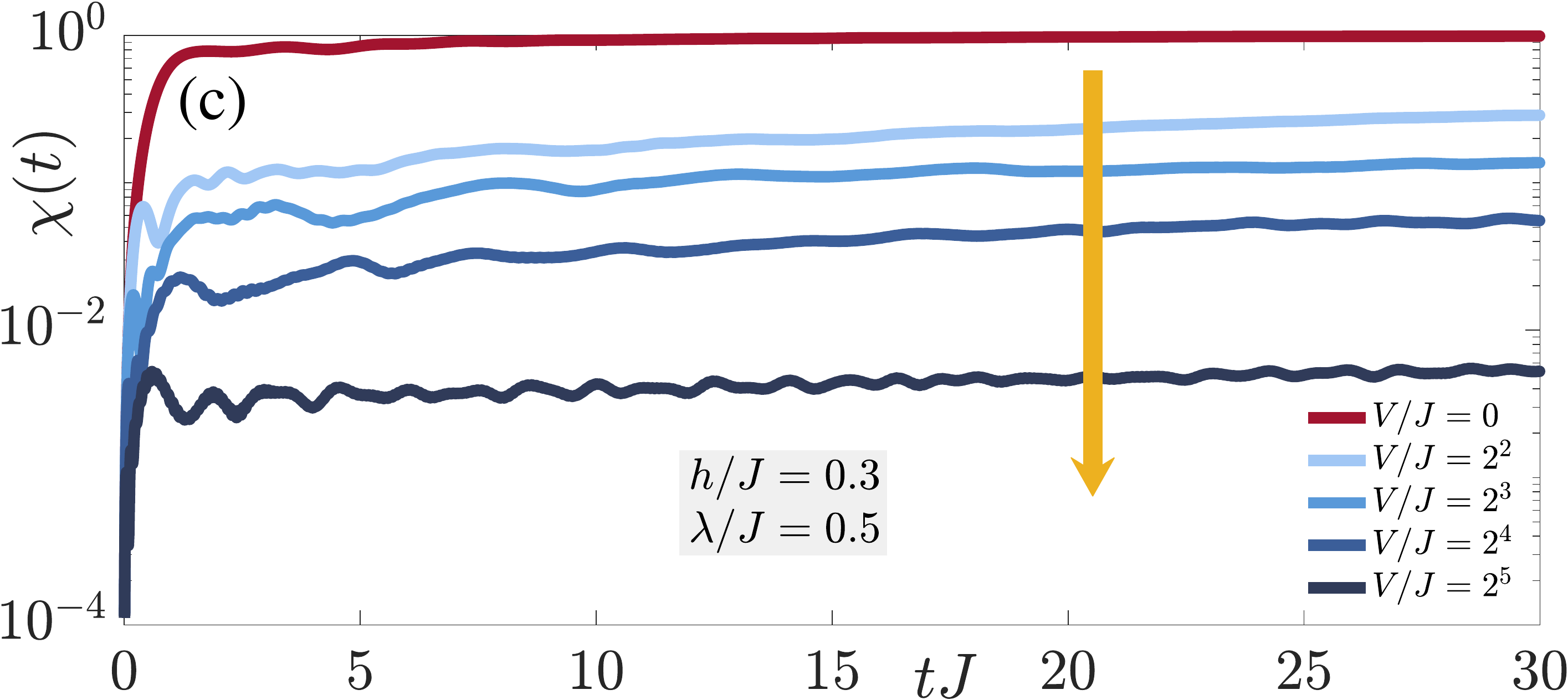}
    \caption{(Color online). Restoring scarred dynamics in the $\mathbb{Z}_2$ LGT in the presence of experimentally relevant errors~\eqref{eq:Z2LGT_H1} by employing the linear gauge protection scheme given in Eq.~\eqref{eq:HW}, based on the principle of the local pseudogenerator given in Eq.~\eqref{eq:LPG}. (a) Fidelity dynamics shows nonthermal revivals in the ideal case (yellow dotted curve in main plot, yellow solid curve in inset), but in the presence of unavoidable errors, the revivals vanish and the behavior of the fidelity is thermal (red dotted curve in inset). Linear gauge protection at moderate values of the protection strength qualitatively restore nonthermal revivals in the fidelity dynamics (different shades of blue), albeit they are quantitatively different from the ideal case due to a renormalization of the gauge theory (see text). (b) The electric flux exhibits persistent oscillations in the ideal case, which vanish in the presence of errors. Linear gauge protection qualitatively restores these persistent oscillations, indicative of robust scarred dynamics. (c) As in the case of the $\mathrm{U}(1)$ QLM in Sec.~\ref{sec:U1QLM}, the robustness of scars in the presence of linear gauge protection is directly connected to the suppression of gauge violations.}
    \label{fig:Z2LGT_FloquetErrors}
\end{figure}

Preparing our system in an initial state $\ket{\psi_0}$ that is an equal superposition of these product states, and quenching with Eq.~\eqref{h_Z2} at $J=1$ and $h=0.3J$, we find that scarring behavior is evident in the dynamics of the fidelity, shown in Fig.~\ref{fig:Z2LGT_FloquetErrors}(a) as a yellow dotted curve in the main plot, and a yellow solid curve in the inset. It is to be noted that the scarring here is not as prominent as in the case of the $\mathrm{U}(1)$ QLM discussed in Sec.~\ref{sec:U1QLM}, and this has already been shown to be the case in Ref.~\cite{Sai2022}. Nevertheless, the behavior of the fidelity shows clear nonthermal revivals, a hallmark of scarred dynamics.

Inspired by a recent implementation of a building block of Eq.~\eqref{h_Z2} in an ultracold-atom experiment \cite{Schweizer2019}, we introduce the gauge-breaking error term
\begin{align}\nonumber
	\lambda\hat{H}_1=\,\lambda\sum_{j=1}^{L}\Big\{&\Big[\hat{b}_j^\dagger\hat{b}_{j+1}\big(\eta_1\hat{\sigma}^+_{j,j+1} +\eta_2\hat{\sigma}^-_{j,j+1}\big)+\mathrm{H.c.}\Big]\\\label{eq:Z2LGT_H1}
	&+\big(\eta_3\hat{n}_j-\eta_4\hat{n}_{j+1}\big)\hat{\sigma}^z_{j,j+1}\Big\},
\end{align}
where the coefficients $\eta_1=0.5110$, $\eta_2=-0.4953$, $\eta_3=0.7696$, and $\eta_4=0.2147$ are determined by an optimal value of a dimensionless driving parameter in the Floquet setup of Ref.~\cite{Schweizer2019}. We emphasize that we are not driving our system, but only performing global quenches. The aim of the Floquet setup employed in the experiment of Ref.~\cite{Schweizer2019} was merely to obtain an effective Floquet Hamiltonian in the form of Eq.~\eqref{h_Z2}, but subleading orders in Floquet theory lead to errors such as those of Eq.~\eqref{eq:Z2LGT_H1}.

Upon quenching by $\hat{H}_0+\lambda\hat{H}_1$, with $\lambda=0.5J$, we see that the fidelity quickly decays and remains vanishing in value up to all accessible evolution times, indicative of thermal behavior, as shown in the inset of Fig.~\ref{fig:Z2LGT_FloquetErrors}(a). This shows the pernicious effect of such errors, even when they are perturbative. It is therefore crucial to protect against such gauge-breaking processes by the introduction of an experimentally feasible technique. However, just by looking at the local generator~\eqref{eq:Z2LGT_Gj}, one notices that it is experimentally quite challenging to implement, being a three-body term. In contrast, we recall that the local generator of the $\mathrm{U}(1)$ QLM is composed of only single-body terms; see Eq.~\eqref{eq:U1QLM_Gj}.

This problem can be resolved by employing a local \textit{pseudogenerator} (LPG) \cite{Halimeh2021stabilizing},
\begin{align}\label{eq:LPG}
    \hat{W}_j=\hat{\sigma}_{j-1,j}^x\hat{\sigma}_{j,j+1}^x+2\hat{n}_j,
\end{align}
which is composed of single and two-body terms, and acts identically to $\hat{G}_j$ of Eq.~\eqref{eq:Z2LGT_Gj} in the target sector, but not necessarily outside of it:
\begin{align}
    \hat{W}_j\ket{\phi}=\ket{\phi}\iff\hat{G}_j\ket{\phi}=\ket{\phi}.
\end{align}
One can then utilize the concept of quantum Zeno subspaces by employing the linear protection scheme \cite{Halimeh2021stabilizing}
\begin{align}\label{eq:HW}
    V\hat{H}_W=V\sum_j\frac{6(-1)^j+5}{11}\hat{W}_j.
\end{align}
At sufficiently large $V$ and when starting in a state in the target sector, quenching with the faulty Hamiltonian $\hat{H}=\hat{H}_0+\lambda\hat{H}_1+V\hat{H}_W$ gives rise to the effective Zeno Hamiltonian of the same form as Eq.~\eqref{eq:QZE}, but with $\hat{H}_0$ given in Eq.~\eqref{h_Z2}, $\lambda\hat{H}_1$ given in Eq.~\eqref{eq:Z2LGT_H1}, and where $\hat{\mathcal{P}}_0$ is the projector onto the target gauge sector. However, unlike the errors~\eqref{eq:U1QLM_H1} in the case of the $\mathrm{U}(1)$ QLM, the errors~\eqref{eq:Z2LGT_H1} do not satisfy $\hat{\mathcal{P}}_0\hat{H}_1\hat{\mathcal{P}}_0=0$, since they also include gauge-invariant processes. The term $\hat{\mathcal{P}}_0\hat{H}_1\hat{\mathcal{P}}_0$ will project out all gauge-breaking processes in $\hat{H}_1$, but keep the gauge-invariant ones in the Zeno Hamiltonian, thereby renormalizing the $\mathbb{Z}_2$ gauge theory. However, we emphasize that these renormalizing terms, being gauge-invariant, preserve the $\mathbb{Z}_2$ gauge symmetry of $\hat{H}_0$, are short-range, and are qualitatively similar to terms found in $\hat{H}_0$.

The effect of this renormalization is seen in the corresponding fidelity dynamics of Fig.~\ref{fig:Z2LGT_FloquetErrors}(a) for a quench under the faulty theory $\hat{H}=\hat{H}_0+\lambda\hat{H}_1+V\hat{H}_W$ (shades of blue). Even though we find that already at small values of $V$ the fidelity again exhibits nonthermal revivals, the latter converge with $V$ to a revival profile quantitatively different from that of the ideal theory, although qualitatively as nonthermal. Nevertheless, it is impressive that scarred dynamics is still robust despite this renormalization, and we clearly see its signature in the fidelity, which is a global quantity quite susceptible to errors.

We now look at the corresponding dynamics of the electric flux, which for the $\mathbb{Z}_2$ LGT is defined as
\begin{align}
    \mathcal{E}(t)=\frac{1}{L-1}\sum_{j=1}^{L-1}\bra{\psi(t)}\hat{\sigma}^x_{j,j+1}\ket{\psi(t)},
\end{align}
with $\ket{\psi(t)}=e^{-i\hat{H}t}\ket{\psi_0}$, and where $\ket{\psi_0}$ is the initial scar state. The result is shown in Fig.~\ref{fig:Z2LGT_FloquetErrors}(b). Whereas in the ideal case we see persistent oscillations up to all accessible evolution times, upon introducing the errors~\eqref{eq:Z2LGT_H1} at $\lambda=0.5J$ without protection, the dynamics equilibrates rather quickly and the oscillations are damped, indicative of expected thermalization. Persistent oscillations are revived upon utilizing the linear protection~\eqref{eq:HW} in the LPG~\eqref{eq:LPG}, which at sufficiently large $V$ converges to a scarred dynamics that is not quantitatively identical to the ideal case, but qualitatively similar and exhibiting persistent oscillations. We note that we do not compute the matter density here since it is always constant at $0.5$ due to the global $\mathrm{U}(1)$ symmetry of the $\mathbb{Z}_2$ LGT.

Finally, we look at the gauge violation~\eqref{eq:viol} with $g_j^\text{tar}=+1,\,\forall j$, in Fig.~\ref{fig:Z2LGT_FloquetErrors}(c). We see a clear connection between a suppressed gauge violation and a more robust scarring dynamics. However, the suppression of the gauge violation does not seem as effective here as in the case of the $\mathrm{U}(1)$ QLM, and we attribute this to the choice of noncompliant sequence $[6(-1)^j+5]/11$ used. The gauge-violation plateau does not settle into a value $\propto\lambda^2/V^2$ for the range of values we consider for the protection strength. However, as we show in the Appendix~\ref{app:supp}, for different newly developed protection schemes, this controlled behavior can also be achieved.

\section{Conclusion}\label{sec:conc}
In summary, we have demonstrated that recently proposed experimentally feasible linear gauge protection schemes render scarred dynamics robust in two paradigmatic systems, the $\mathrm{U}(1)$ quantum link model and the $\mathbb{Z}_2$ lattice gauge theory. This is remarkable because fine-tuned phenomena such as quantum many-body scars require persistent revivals in the fidelity itself, which indicate that the wave function is localized in a cold subspace containing the initial state. Such a global quantity as the fidelity is more susceptible to errors, but our numerical simulations show that it is nevertheless well-restored through the considered gauge protection schemes. We note that the $\mathbb{Z}_2$ scars are reproduced qualitatively, not quantitatively, in the presence of certain experimentally relevant errors that renormalize the gauge theory, showing the generality of our approach in stabilizing a unique physical phenomenon.

In addition to the fidelity, we have also computed local observables such as the electric flux and the matter density (chiral condensate), which under ideal dynamics exhibit persistent oscillations up to all accessible evolution times. Even though unprotected errors wash out these oscillations, linear gauge protection at experimentally feasible values of the protection strength can restore these persistent oscillations, and in some cases does so quantitatively relative to the ideal case.

We have also computed the gauge violation dynamics, which demonstrate a direct connection with the restored scarring behavior. The more gauge violations are suppressed, the better restored is the scarring dynamics. This implies that leakage out of the target gauge sector exposes its quantum many-body scars to other subspaces in the total Hilbert space that couple nontrivially with the scars, leading to thermal behavior when no protection is employed. We have explained our findings with the concept of quantum Zeno subspaces, through which the target sector is energetically isolated from other gauge sectors up to times at least linear in the protection strength. As our results demonstrate, scarring dynamics is robust under linear gauge protection at moderate values of the protection strength up to timescales that are relevant to modern synthetic quantum matter implementations of gauge theories. 

Given the experimental feasibility of the linear gauge protection schemes that we have employed in this work, we expect our results to be directly implementable in modern quantum-simulation realizations of lattice gauge theories, including ultracold-atom optical lattices and Rydberg-atom setups with optical tweezers. Furthermore, we expect our linear gauge protection schemes to apply equally well to recently discovered quantum many-body scars in the spin-$S$ $\mathrm{U}(1)$ quantum link model for $S>1/2$ \cite{Desaules2022a}.

\begin{acknowledgments}
J.C.H.~is grateful to Jean-Yves Desaules, Ana Hudomal, Zlatko Papi\'c, and Guoxian Su for work on related topics. L. B. acknowledges Utso Bhattacharya, Daniel Gonz\'alez-Cuadra, Maciej Lewenstein, Adith Sai Aramthottil and Jakub Zakrzewski for discussions on related topics. This project has received funding from the European Research Council (ERC) under the European Union’s Horizon 2020 research and innovation programm (Grant Agreement no 948141) — ERC Starting Grant SimUcQuam, and by the Deutsche Forschungsgemeinschaft (DFG, German Research Foundation) under Germany's Excellence Strategy -- EXC-2111 -- 390814868, and from the NSF through a grant for the Institute for Theoretical Atomic, Molecular, and Optical Physics at Harvard University and the Smithsonian Astrophysical Observatory. This work is part of and supported by Provincia Autonoma di Trento, the ERC Starting Grant StrEnQTh (project ID 804305), the Google Research Scholar Award ProGauge, and Q@TN — Quantum Science and Technology in Trento.
\end{acknowledgments}

\appendix

\begin{figure}[t!]
    \centering
    \includegraphics[width=\columnwidth]{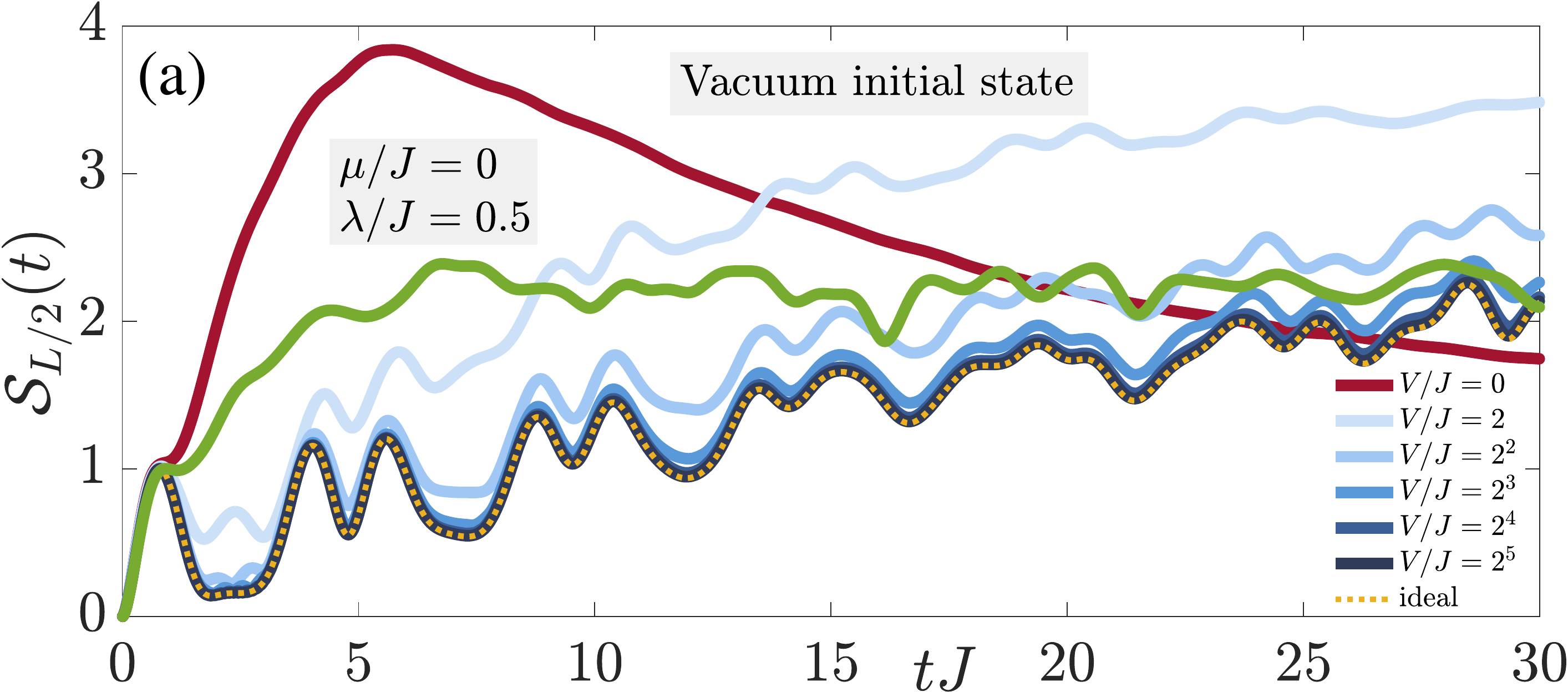}\\
	\vspace{1.1mm}
    \includegraphics[width=\columnwidth]{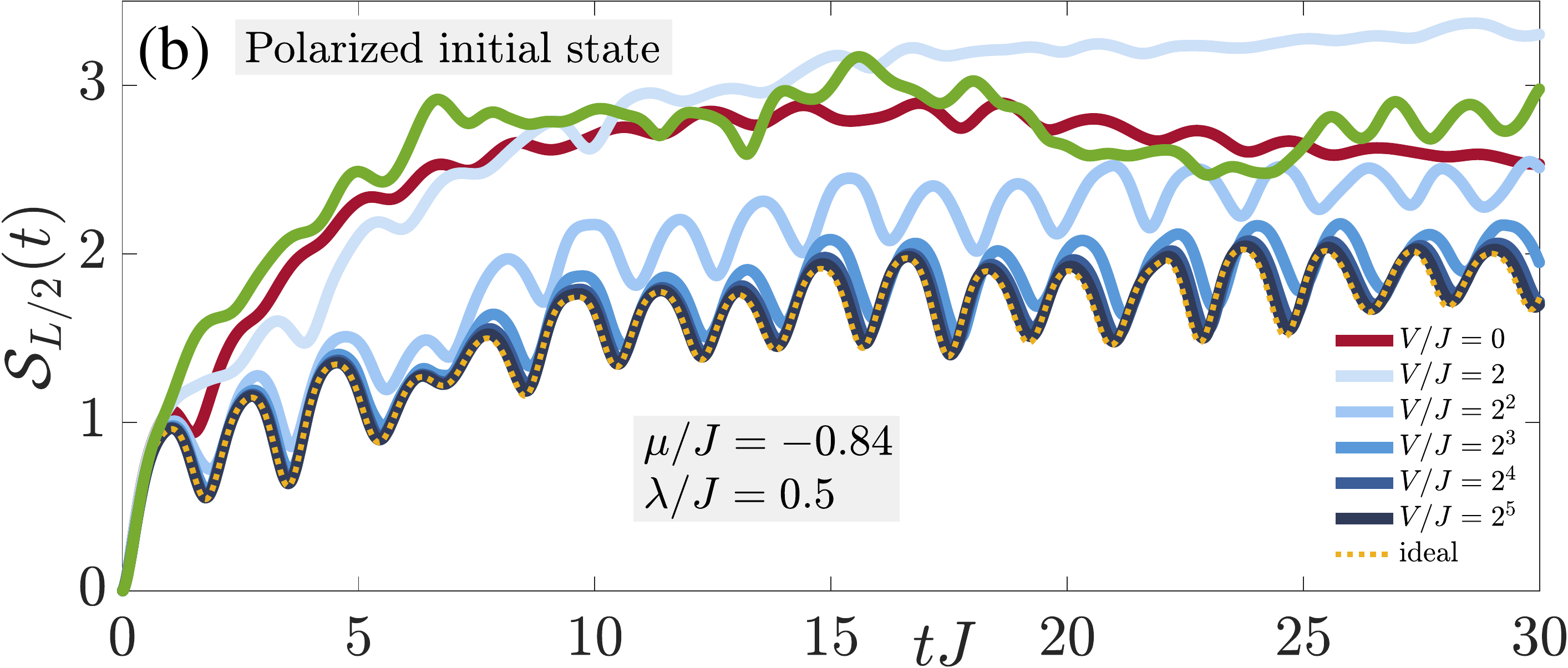}
    \caption{(Color online). Dynamics of the mid-chain entanglement entropy in the $\mathrm{U}(1)$ QLM. (a) Quench of the vacuum state with $\hat{H}=\hat{H}_0+\lambda\hat{H}_1+V\hat{H}_G$ at $\mu=0$, with $\hat{H}_0$, $\lambda\hat{H}_1$, and $V\hat{H}_G$ given in Eqs.~\eqref{eq:U1QLM_H0},~\eqref{eq:U1QLM_H1}, and~\eqref{eq:HG}, respectively. For reference, we include the result for the quench of the charge-proliferated (polarized) state with $\hat{H}_0$ at $\mu=0$ (green solid curve). (b) Quench of the polarized state with $\hat{H}=\hat{H}_0+\lambda\hat{H}_1+V\hat{H}_G$ at $\mu=-0.84J$, and that of the vacuum with $\hat{H}_0$ at $\mu=-0.84J$. In both cases, linear gauge protection at sufficiently large $V$ reproduces the localized scarred dynamics of the ideal case (yellow dotted curve), which exhibits an anomalously low and slowly growing mid-chain entanglement entropy over the considered time-evolution window. In our numerics, we have employed open boundary conditions with $L_\text{m}=L=10$ matter sites and $L_\text{g}=L-1=9$ gauge links.}
    \label{fig:u1qlm_EE}
\end{figure}

\begin{figure}[t!]
    \centering
    \includegraphics[width=\columnwidth]{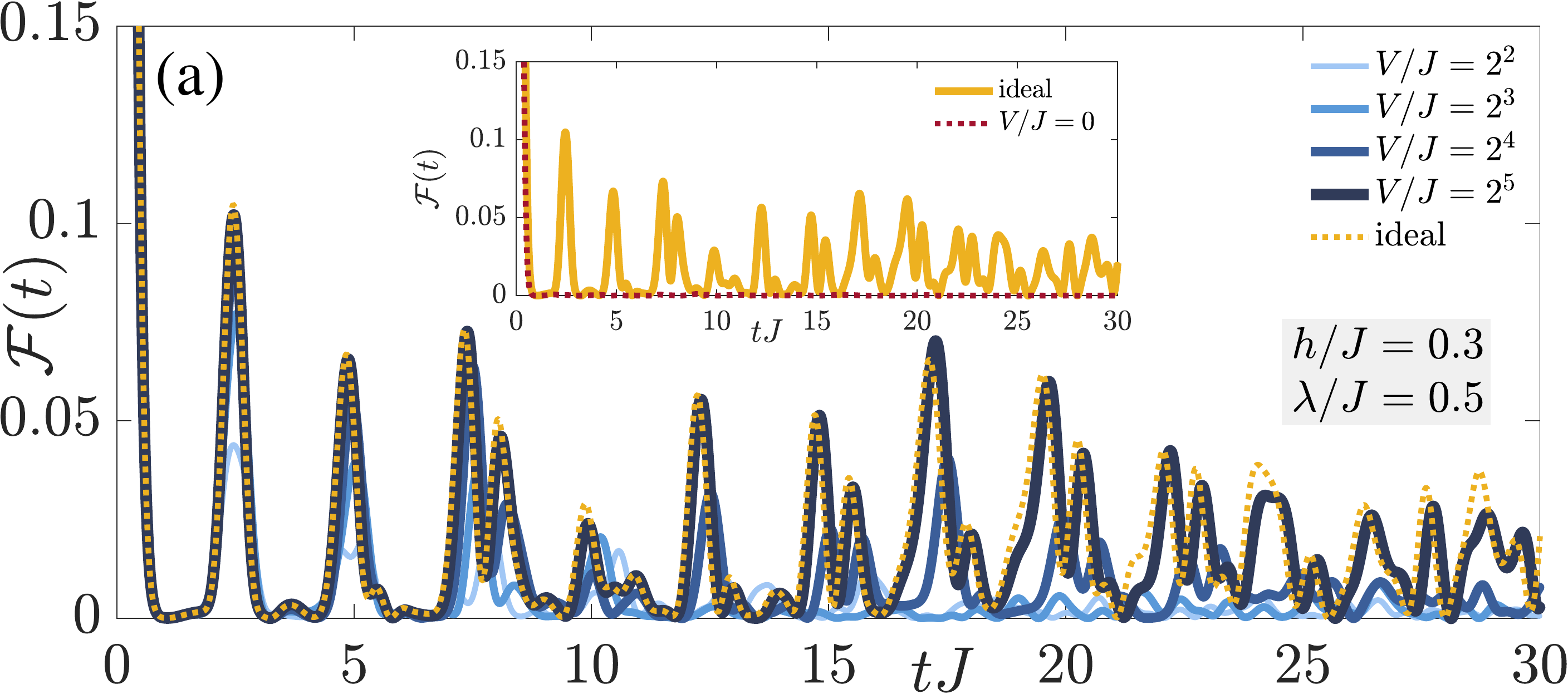}\\
	\vspace{1.1mm}
    \includegraphics[width=\columnwidth]{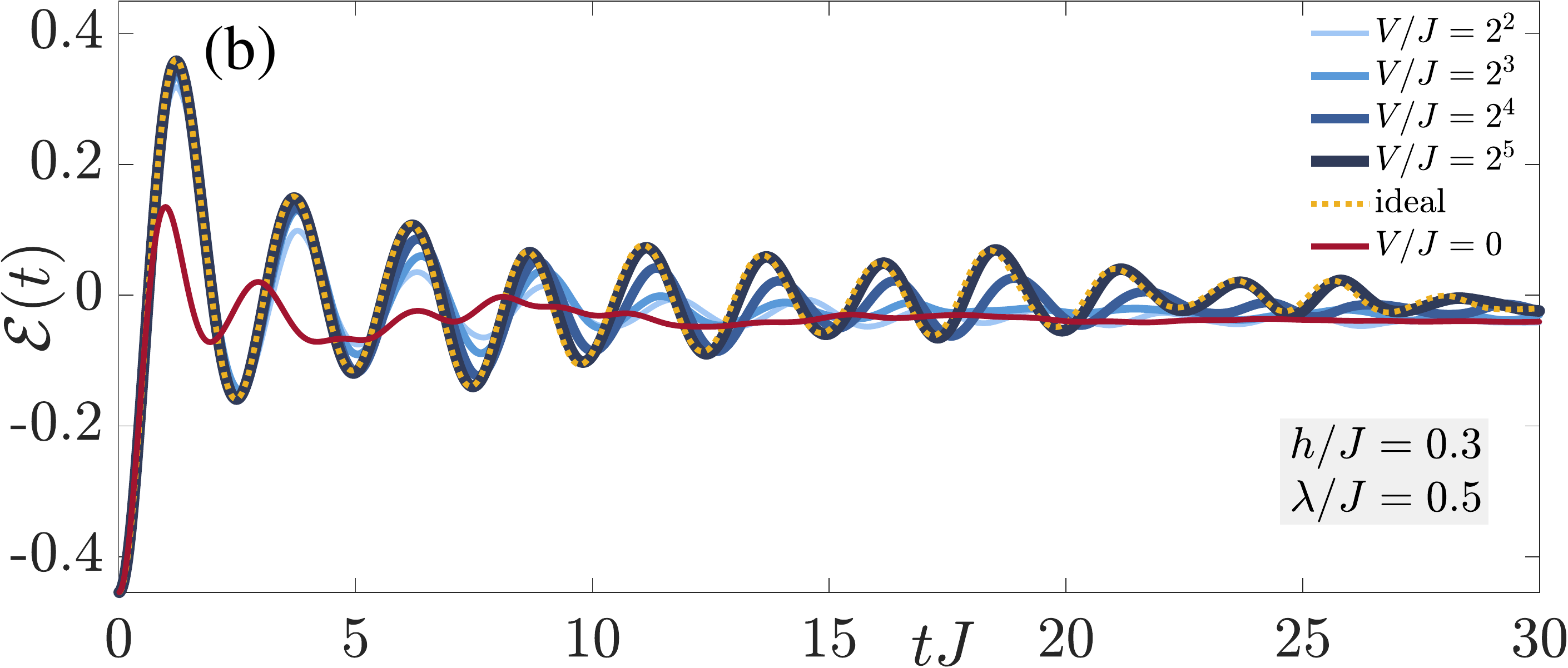}\\
	\vspace{1.1mm}
    \includegraphics[width=\columnwidth]{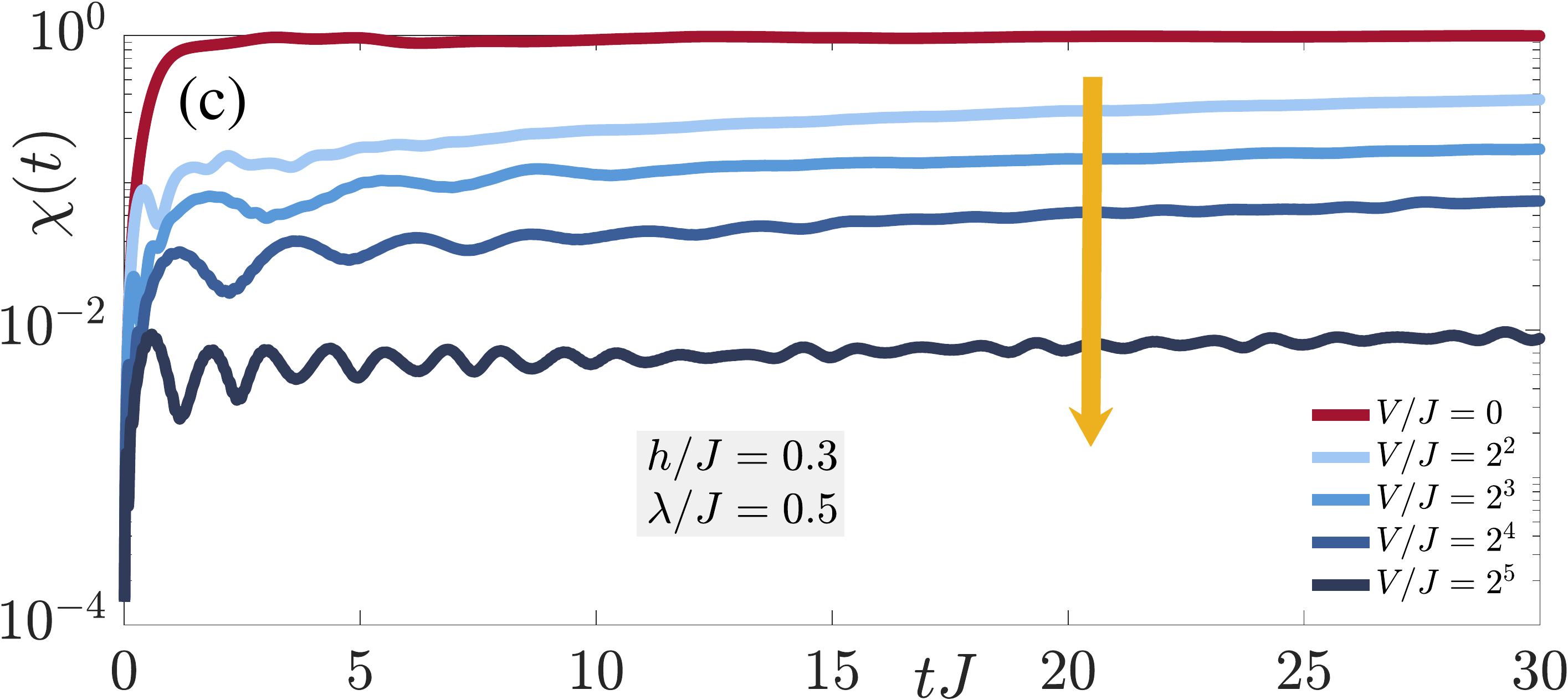}
    \caption{(Color online). Same as Fig.~\ref{fig:Z2LGT_FloquetErrors} but with the errors of Eq.~\eqref{eq:SimpleError} instead of those in Eq.~\eqref{eq:Z2LGT_H1}. The qualitative picture is the same, but we further achieve much better quantitative agreement with the ideal case at larger values of $V$, because now $\hat{\mathcal{P}}_0\hat{H}_1\hat{\mathcal{P}}_0=0$.}
    \label{fig:Z2LGT_SimpleError}
\end{figure}

\begin{figure}[t!]
    \centering
    \includegraphics[width=\columnwidth]{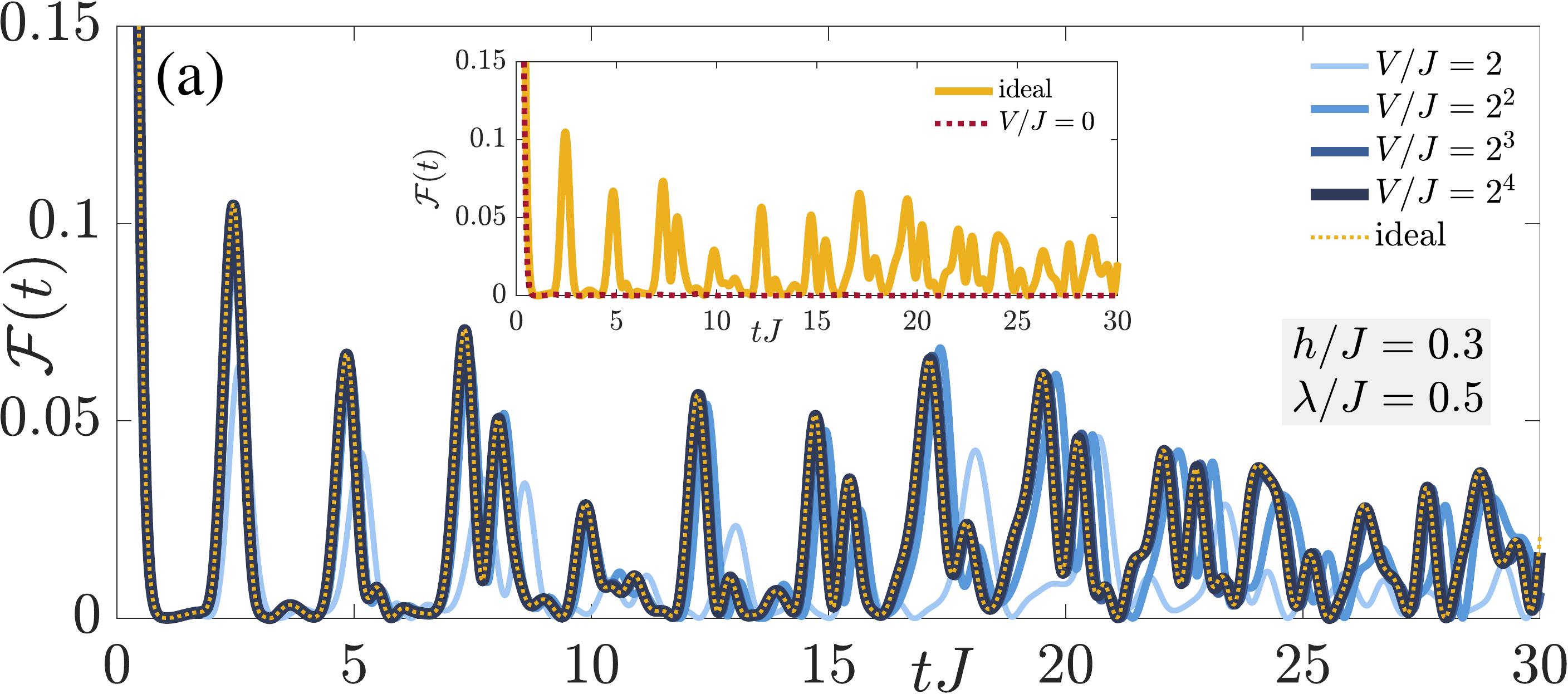}\\
	\vspace{1.1mm}
    \includegraphics[width=\columnwidth]{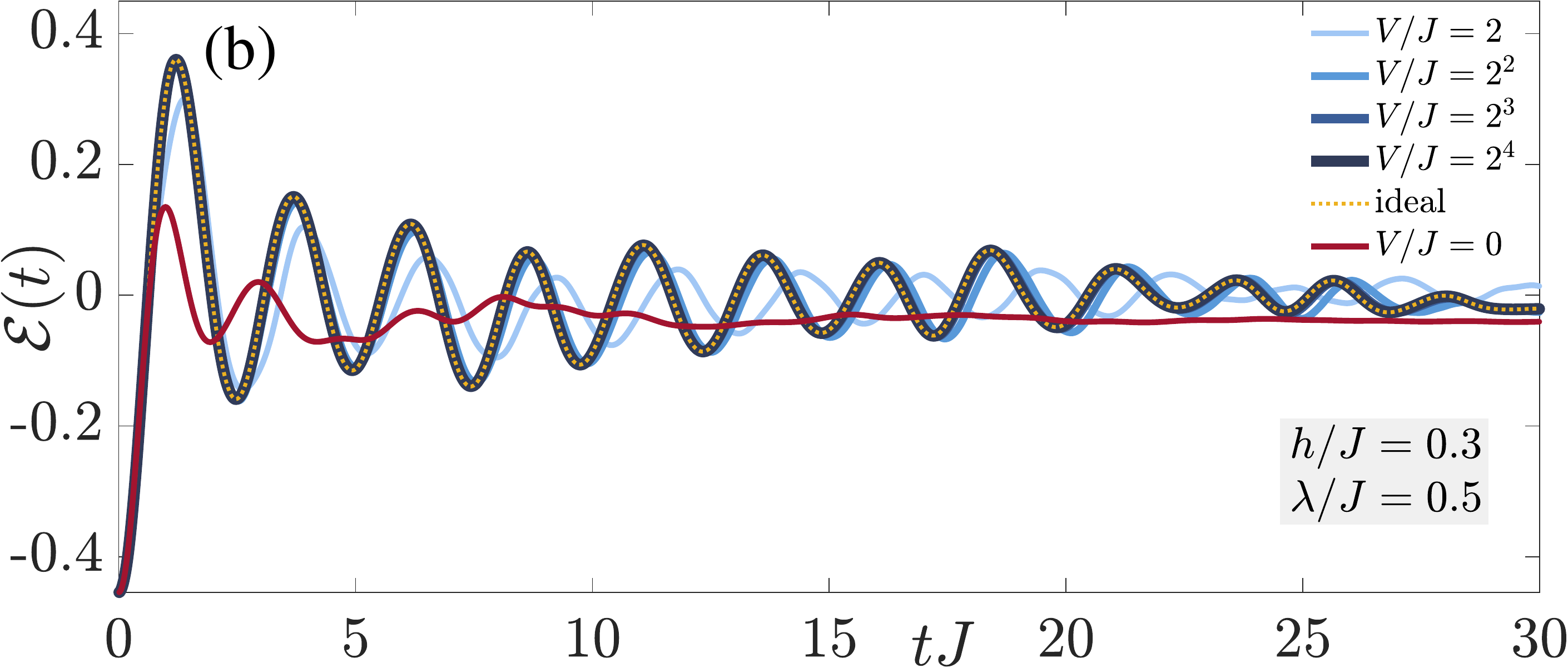}\\
	\vspace{1.1mm}
    \includegraphics[width=\columnwidth]{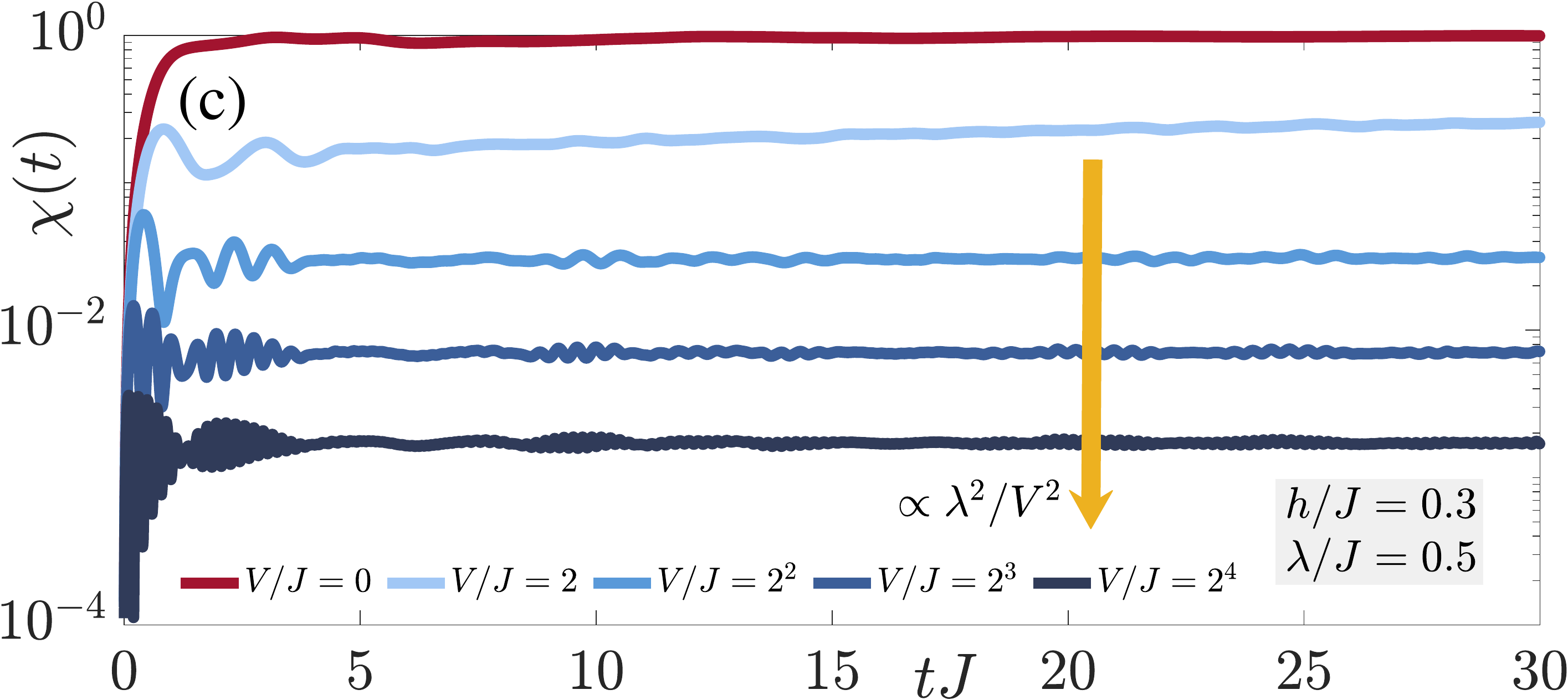}
    \caption{(Color online). Same as Fig.~\ref{fig:Z2LGT_SimpleError}, but employing the Stark gauge protection term~\eqref{eq:SGP} instead of Eq.~\eqref{eq:HW}. The quantitative agreement with the ideal case is excellent for both (a) the fidelity and (b) the electric flux, but (c) additionally we now see a controlled gauge violation suppressed $\propto\lambda^2/V^2$.}
    \label{fig:Z2LGT_SimpleError_Stark}
\end{figure}

\section{Supporting results}\label{app:supp}
Here, we provide supplemental numerical results in support of the main conclusions of our work.

\subsection{Mid-chain entanglement entropy}
We now consider the behavior of the mid-chain entanglement entropy $\mathcal{S}_{L/2}(t)$, shown in Fig.~\ref{fig:u1qlm_EE}(a,b) for the cases of resonant and detuned scarring, respectively, in the $\mathrm{U}(1)$ QLM~\eqref{eq:U1QLM_H0} in the presence of experimentally relevant errors~\eqref{eq:U1QLM_H1} and with linear gauge protection~\eqref{eq:HG}.

In the ideal case ($\lambda=V=0$), we see that the scarred dynamics (yellow dotted curve) exhibit anomalously low mid-chain entanglement entropy with significantly slower growth compared to the case of the same quench but starting in a nonscar initial state (green solid curve) such as the charge-proliferated state in Fig.~\ref{fig:u1qlm_EE}(a) and the vacuum state in Fig.~\ref{fig:u1qlm_EE}(b).

Upon introducing errors ($\lambda=0.5J$) and without protection ($V=0$), we see a fast growth in the mid-chain entanglement entropy even when starting in the correct initial scar state. However, upon turning on the linear gauge protection, we find that the entanglement entropy is suppressed, and its growth is slowed down, indicating that scars are trapping the dynamics in a low-entropy subspace. At a moderate value $V=16J$ of the protection strength, we already find excellent quantitative agreement with the ideal case for all evolution times in Fig.~\ref{fig:u1qlm_EE}.

We note that due to numerical overhead, the results of Fig.~\ref{fig:u1qlm_EE} are computed for an open chain with $L_\text{m}=L=10$ matter sites and $L_\text{g}=L-1=9$ gauge links with open boundary conditions.

\subsection{Different kinds of gauge-breaking errors}
In Sec.~\ref{sec:Z2LGT}, we have considered the errors~\eqref{eq:Z2LGT_H1}, which are inspired from a recent ultracold-atom experiment based on a Floquet setup used to engineer a building block of a $\mathbb{Z}_2$ LGT~\cite{Schweizer2019}. As we have explained, these errors contain also gauge-invariant processes, which renormalize the emergent gauge theory due to quantum Zeno dynamics up to timescales at least linear in $V$. However, in other setups different errors may arise. Let us now choose a different error of the form
\begin{align}\label{eq:SimpleError}
    \lambda\hat{H}_1=\lambda\sum_{j=1}^{L-1}\big(\hat{b}^\dagger_j\hat{b}_{j+1}+\hat{b}_j\hat{b}_{j+1}^\dagger+\hat{\sigma}^z_{j,j+1}\big).
\end{align}
This corresponds to the gauge-noninvariant processes of matter annihilation or creation without the simultaneous change in the electric-field configuration required to satisfy Gauss's law, and vice versa. In particular, this error satisfies $\hat{\mathcal{P}}_0\hat{H}_1\hat{\mathcal{P}}_0=0$, meaning that the emergent Zeno Hamiltonian is the same as the ideal theory up to an error of the order $\mathcal{O}(tV_0^2L^2/V)$. We now repeat the quench results of Fig.~\ref{fig:Z2LGT_FloquetErrors} but with the error of Eq.~\eqref{eq:SimpleError}, with the corresponding results provided in Fig.~\ref{fig:Z2LGT_SimpleError}. We find that in this case the quantitative agreement of the fidelity and electric-flux dynamics at sufficiently large $V$ with those of the ideal case is much better than in Fig.~\ref{fig:Z2LGT_FloquetErrors}. However, the gauge violation still does not show a controlled suppression $\propto\lambda^2/V^2$, and this may be an issue with the protection scheme~\eqref{eq:HW} itself. Indeed, there are various ways of choosing the LPG coefficients in such linear gauge protection terms, and different coefficients will achieve a better partition of the gauge sectors into quantum Zeno subspaces \cite{Halimeh2021stabilizingDFL}. In the following, we shall explore a recently introduced protection scheme that solves the problem and leads to a controlled gauge violation at sufficiently large yet experimentally feasible values of $V$.

\subsection{Stark gauge protection}
Recently, Stark gauge protection (SGP),
\begin{align}\label{eq:SGP}
    V\hat{H}_\text{SGP}=V\sum_jj\hat{W}_j,
\end{align}
has been introduced, showing superior performance compared to Eq.~\eqref{eq:HW} in stabilizing and enhancing disorder-free localization in the $\mathrm{U}(1)$ QLM and $\mathbb{Z}_2$ LGT \cite{lang2022disorder}. Even though the states considered in Ref.~\cite{lang2022disorder} are superpositions over an extensive number of gauge sectors rather than fine-tuned scar states in a target gauge sector, it will be interesting to investigate the power of SGP in protecting scarred dynamics. We therefore repeat the quench of Fig.~\ref{fig:Z2LGT_SimpleError} but with SGP instead of Eq.~\eqref{eq:HW}, and present the corresponding results in Fig.~\ref{fig:Z2LGT_SimpleError_Stark}. The quantitative agreement in the fidelity and electric-flux dynamics with those of the ideal theory at a given value of $V$ is significantly better under SGP compared to Eq.~\eqref{eq:HW}, albeit the qualitative performance is the same. However, the dynamics of the gauge violation is qualitatively far better compared to Eq.~\eqref{eq:HW}, where under SGP we see a controlled plateau $\propto\lambda^2/V^2$ at sufficiently large values of $V$.

\bibliographystyle{quantum}
\bibliography{Robustness_biblio}

\end{document}